\documentstyle[12pt,aaspp]{article}

\begin{document}

\title{The Norris Survey of the Corona Borealis Supercluster: \\ II.
Galaxy Evolution with Redshift and Environment}

\author{Todd A. Small\altaffilmark{1} and Wallace L.W. Sargent}
\affil{Palomar Observatory, California Institute of Technology,
Pasadena, CA 91125 \\
Electronic mail: tas@ast.cam.ac.uk, wws@astro.caltech.edu}

\author{Donald Hamilton}
\affil{Max-Planck-Institut f\"ur Astronomie, K\"onigstuhl 17,
Heidelberg  D-69117, Germany \\
Electronic mail: hamilton@mpia-hd.mpg.de}

\altaffiltext{1}{present address:  Institute of Astronomy,
University of Cambridge, Madingley Road, Cambridge CB3 0HA, UK}

\begin{abstract}

We measure the field galaxy luminosity function (LF) as a function
of color and redshift from $z = 0$ to $z = 0.5$ using galaxies from
the Norris Survey of the Corona Borealis Supercluster.  The data set
consists of 603 field galaxies with $0 < z \le 0.5$ and spans
a wide range in apparent magnitude ($14.0^m \lesssim r \lesssim 21.5^m$),
although our field galaxy LF analysis is limited to 493 galaxies
with $r \le 20.0^m$.
We use the observed $g-r$ colors of the galaxies to compute accurate
corrections to the rest $B_{AB}$ and $r$ bands.  
We find that our local $r$-band LF, when normalized to counts in
high galactic latitude fields, agrees well with the local LF measured
in the Las Campanas Redshift Survey.  Our $B_{AB}$-band local LF,
however, does not match the $b_j$-band LF from the Stromlo/APM survey,
having a normalization 1.6 times higher.
We see compelling evidence that the $B_{AB}$-band field galaxy
LF evolves with redshift.  The evolution is strongest
for the population of star-forming galaxies with [\ion{O}{2}] $\lambda$3727 
rest-frame
equivalent widths greater than 10\AA.  The population of red, quiescent
galaxies shows no sign of evolution to $z = 0.5$.
The evolution of the LF which we observe is consistent with the findings of
other faint galaxy redshift surveys.  The fraction of galaxies with
[\ion{O}{2}] emission increases rapidly with redshift, but the fraction
of galaxies with strong H$\delta$ absorption, a signature of a burst
of star-formation, does not.  We thus conclude that the star formation
in distant galaxies is primarily long-lived.

We also compute the LFs of the Corona Borealis
supercluster ($z \approx 0.07$, 419 galaxies with $14.1^m \le r \le 20.0^m$) 
and the Abell 2069 supercluster ($z \approx 0.11$, 318 galaxies with
$15.1^m \le r \le 20.0^m$).  The shapes of the two supercluster luminosity
functions are broadly similar to the shape of the local luminosity
function.  However, there are important differences.  Both supercluster
LFs have an excess of very bright galaxies.  In addition, the
characteristic magnitude of the Corona Borealis supercluster LF is
roughly half a magnitude brighter than that of the local field galaxy
LF, and there is a suggestion of an upturn in the LF for galaxies
fainter than $M(B_{AB}) \sim -17^m$.

\end{abstract}


\section{Introduction}

With the introduction of efficient multi-object spectrographs
on 4m-class telescopes, it has become possible to construct
large samples of faint galaxies with measured redshifts.  With such
a sample, one can compute the luminosity function (LF) of galaxies as
a function of redshift and thereby directly observe the evolution
(or lack thereof) of the galaxy population.  Several groups have
now presented the results of deep, faint galaxy redshift surveys
(\cite{Lilly:lf}, CFRS; \cite{Ellis:lf}, Autofib; \cite{Cowie:lf}; \cite{Lin:cnoc}, CNOC).
The conclusions from these surveys are in broad agreement:  the population
of blue,
star-forming galaxies has evolved strongly since $z \sim 0.5$ while
the population of red galaxies shows at most modest signs of evolution (although,
see Kauffmann, Charlot, \& White (1996) for an alternative analysis of
the red galaxies).
However, there are important differences as well.  Lin et al. (1996a)
demonstrate that the LFs from the various groups
are formally inconsistent with each other.  Since there are many
selection effects involved with the construction and analysis of faint
redshift surveys, it is difficult to pinpoint the reasons for the
disagreement between the various groups.  While it is likely that
the small numbers of galaxies in each survey and the small areas covered
are partly responsible, it is also likely that systematic errors 
are in important contributor to the differences in detail.

Quantitative estimates of the evolution are, of course,
dependent upon having a reliable measurement of the local LF,
and it is, therefore, of concern that there remain
considerable uncertainties about the {\it local} LF.
The LFs derived from large-area local
redshifts survey (e.g., the Stromlo/APM survey, Loveday et al. 1992;
the CfA survey, Marzke, Huchra, \& Geller 1994a; the Las Campanas Redshift Survey,
Lin et al. 1996b) all have similar shapes,
but there are still substantial differences over the overall normalization,
the characteristic luminosity,
and the slope at low luminosities.  The rapid evolution at $z \sim 0.1$
required to match steep $B$-band counts at intermediate magnitudes
$17^m < B < 21^m$ (Maddox et al. 1990) could be reduced if the normalization
or the faint-end slope have been underestimated.  The results of the
largest of the local surveys, the Las Campanas Redshift Survey (LCRS)
with 18678 galaxies used in the LF analysis and a median
redshift of $z \approx 0.1$, are
seemingly consistent with both a low normalization and a flat faint-end
slope.  The LCRS is selected from CCD drift scans rather than
photographic plates and surveys what should be a fair volume of the universe
(Shectman et al. 1996, Davis 1996).  It also probes both the Southern
and Northern Galactic Caps.  Accordingly, the local luminosity
function computed from their data should be free from
systematic photometric errors and fluctuations in large-scale
structure in the distribution of galaxies.  However, both the
CfA survey and the Autofib survey find a normalization
which is a factor of 2 higher than that obtained from the LCRS.
While the normalization of the CfA
survey can be questioned on the grounds that it does not sample a
fair volume, the Autofib survey is the concatenation of many fields
distributed across the sky.
The Autofib survey is particularly important
because the galaxy sample was selected with a much fainter surface
brightness threshold than any of the other local surveys.
McGaugh (1994) emphasizes
that a large population of intrinsically luminous but low surface
brightness galaxies may be missed in the shallow photometry on which all
the local surveys, except Autofib, are based.

A steep faint-end slope of the LF, with a power
law exponent of $\alpha \lesssim -1.5$, is a natural prediction
of galaxy formation theories based on hierarchical structure formation
models (Kauffmann, Guiderdoni, \& White 1994).  There is only weak evidence for
a steep faint-end slope in the local field galaxy LF.
Marzke et al. (1994b) report an upturn in the luminosity
function of late-type galaxies with $M(B) > -16 + \log_{10}h$,
but LCRS, Autofib, and Cowie et al. (1996) all
derive a flat faint-end slope.  There is, however, evidence for
a steep faint-end slope in galaxy clusters (e.g., De Propris et al. 1995,
Bernstein et al. 1995).  Environmental influences on galaxy
evolution may be reflected in variations of the LF
for galaxies in different environments, and it is therefore important
to measure the LF in a variety of environments.

In this paper, we investigate the evolution and environmental
dependence of the galaxy LF based on data obtained during
the course of our
redshift survey of the Corona Borealis supercluster.  The primary motivation
for the survey was to study the dynamics of the supercluster.  However,
the majority of galaxies for which we measured redshifts actually lie
behind the Corona Borealis supercluster, thus providing a sample suitable
for study of the evolution of the LF.  The galaxies
were originally selected from plates taken as part of the Second
Palomar Observatory Sky Survey (POSS-II; \cite{Reid:poss2}) and have
been calibrated in the Gunn $g$ and $r$ bands, which correspond roughly
to the photographic $J$ and $F$ bands.  Previous redshift surveys have
generally
been either selected in bluer bands ($B$), for sensitivity to changes
in star-formation rates, or redder bands ($I$ and $K$), for sensitivity
to old stellar populations which more reliably trace stellar mass.  Although we
had no option but to use the $g$ and $r$ bands, the two bands 
turn out fortuitously to have the
virtue that corrections to the rest $B$ band, where
LFs are traditionally computed and compared, are small
since the
$g$ band matches the rest $B$ band at $z \approx 0.2$ and the $r$ band
matches the rest $B$ band at $z \approx 0.5$.
The CNOC survey used photometry in $g$ and $r$ as well, and so it is
particularly interesting to compare our results to that survey since
there should be no systematic effects due to using different passbands
for galaxy selection.  Finally, with over 400 redshifts in the
Corona Borealis supercluster, and roughly 300 in a background
supercluster, we can explore the variation of the LF
from the field to the supercluster environment.

The paper, the second in the series presenting results from the
Norris Survey of the Corona Borealis Supercluster, is organized as follows.  In \S 2, we summarize
our survey, particularly emphasizing those features that are directly
relevant to the computation of the LF.  We discuss
the details of the computation of the LF in \S 3.
The results are given in \S 4 for both field galaxies and for the
two superclusters individually and are discussed in \S 5. 
Finally, we summarize our conclusions in \S 6.

We use a Hubble constant $H_0 = 100h$ km s$^{-1}$ Mpc$^{-3}$
and a deceleration parameter $q_0 = 0.5$.  For comparison to the
most recent work in the field (e.g., CFRS and CNOC), we use the
AB-normalized $B$ band, $B_{AB}$ (Oke 1974).  The offsets from
$B_{AB}$ to $b_j$ and $B$ are $b_j \approx B_{AB}$ and
$B = B_{AB} - 0.14^m$ (Fukugita, Shimasaku, \& Ichikawa 1995).

\section{The Norris Survey of the Corona Borealis Supercluster}

The Norris Survey of the Corona Borealis Supercluster has been described
in detail in Small et al. (\nocite{Small:data}1996), Paper I
of the current series, and will be only briefly
reviewed here.  The core of the supercluster covers a $6\deg \times 6\deg$
region of the sky centered at right ascension $15^h20^m$, declination
$+30\deg$ and consists of 7 rich Abell clusters at $z \approx 0.07$.
Since the field-of-view of the 176-fiber Norris Spectrograph is
only 400 arcmin$^2$, we planned to observe 36 fields arranged in a
rectangular grid with a grid spacing of 1\deg.  As it turned out,
we successfully observed 23 of the fields and 9 additional fields
along the ridge of galaxies between Abell 2061 and Abell 2067, yielding
redshifts for 1491 extragalactic objects.  We have extended our survey with
163 redshifts from the literature, resulting in 1654 redshifts in the
entire survey.  1022 of these galaxies lie beyond the Corona Borealis
Supercluster, although of these 1022, 325 (318 with $r \le 20.0^m$)
galaxies are in a background
supercluster ($z \approx 0.11$) which we have dubbed the
``Abell 2069 Supercluster.''  The survey fields are distributed across
an area of 25 deg$^2$.
The total area covered by the 32 observed fields, albeit sparsely
sampled, is 2.99 deg$^2$.

As noted above and described in detail in Paper I, the objects have been selected from POSS-II photographic plates of POSS-II field 449, which
neatly covers the entire core of the supercluster.
We have both a $J$ (Kodak III-aJ
emulsion with a GG395 filter) plate and an $F$ (Kodak III-aF emulsion with a
RG610 filter) plate.
The plates were digitized
with 1 arcsec$^2$ pixels at the Space Telescope Science Institute and
then processed using the Sky Image Cataloging and Analysis Tool (SKICAT,
\cite{Weir:phd}).  The instrumental intensities recorded by SKICAT
were calibrated with CCD sequences in the $g$ and $r$ bands of galaxies in
Abell 2069.  The random magnitude errors are 0.25$^m$
for $g$ and $r$ brighter than 21$^m$ and become substantially worse at
fainter magnitudes. (We describe how we correct our computed LFs
for these magnitude errors in \S 3.2.) 
With the SKICAT system, the star-galaxy
separation is 90\% accurate to $r \approx 20^m$.  Our LF
analysis is limited to galaxies with $r \le 20^m$. 

Since our original motivation for the survey was to study the dynamics
of the Corona Borealis supercluster, we chose a comparatively
high spectral resolution for a faint galaxy redshift survey.
A third of the objects were observed with $\sim 8$\AA\
spectral resolution during the period when the largest CCD available at
Palomar was a
1024$^2$ device with 24\micron\ pixels; the rest were observed with $\sim 4$\AA\ resolution with a very efficient 2048$^2$ CCD (also with 24\micron\ 
pixels).  Since the operation of retrieving the fibers for one set-up
and redeploying the fibers for another takes roughly an hour with the Norris
Spectrograph,
we decided to observe only two fields per night in order to minimize
the amount of time lost due to changing fields.  Thus, our exposures
were 2-4 hours long, and we generally obtained high quality spectra on
$r \sim 20^m$ galaxies.

In Figure \ref{figures:success}, we plot our success rate, the fraction of
objects  (i.e., including stars and quasars) on which fibers were deployed
for which we successfully measured redshifts, as a function of magnitude.
Figure \ref{figures:success} shows that our success rate falls
substantially below unity beyond $r = 18.5^m$.
Therefore, we have computed weights
for each galaxy to correct for a our incomplete sampling.  The weight
for a particular object is defined to be simply the ratio of the total
number of objects in the photometric catalog
to the number of objects with redshifts in an
interval of $0.5^m$ centered on the magnitude
of the object.  This prescription assumes that the redshift distribution
of the objects
for which we failed to measure redshifts is identical to the redshift
distribution of the objects for which we successfully measured redshifts.
The color distribution
of the objects that we observed but failed to identify is similar
to that of the objects that we successfully observed,
leading us to conclude that we do not suffer biases against
particular types of galaxies (Paper I).  
Moreover, we do not believe that we should have redshift-dependent biases
in our success rate.  Since we
limit the computation of the LF to $z < 0.5$, we
are unlikely to be affected by a bias in redshift.  The 4000\AA\ break
and Ca H, Ca K lines of old stellar populations and the [\ion{O}{2}] line
of star-forming galaxies are all within our spectral range out to $z = 0.5$.
We plot the calculated weights as a function of $r$ magnitude for all galaxies
with $r \le 20^m$ and which satisfy our surface brightness threshold (see
below) in Figure \ref{figures:weights}.  
The weights are greater than unity even for bright
galaxies because of the sparse sampling of our survey area.  (Since
the very brightest galaxies ($r < 15^m$) produce scattered light contamination of nearby spectra
on the CCD, we usually did not place fibers on galaxies with $r < 15^m$,
and thus the weights increase for the very brightest galaxies.)

In Paper I, we carefully studied the surface
brightness selection effects present in our sample.  We found that
by restricting our sample to objects with $r \le 20^m$ and with core magnitudes
$r_{\rm core} \le 21.7^m$ (where the core magnitude is the integrated
magnitude within the central 9 arcsec$^2$),
we are free from surface brightness selection effects.  For comparison,
$r_{\rm core} = 21.7^m$ corresponds to a central surface brightness of
$\mu = 24.1$ $r$ mag arcsec$^{-2}$ for a galaxy with an $L^\ast$
($\approx -20.3+5\log_{10}h$ mag in the $r$ band) disk.

\section{Calculation of the Luminosity Function}

\subsection{$k$-Corrections}

We compute galaxy LFs in the rest-frame
$B_{AB}$ band and, for the local LF, in the rest-frame
Gunn $r$ band as well.
Rest-frame colors and $k$-corrections
are computed from the spectral energy distributions compiled
by Coleman, Wu, \& Weedman (1980, hereafter CCW).  We assign
each galaxy a spectral type based on its $g - r$ color and its
redshift.  Following Lilly et al. (1995), the spectral type is
a real number which takes the values 0 for an elliptical galaxy,
2 for an Sbc galaxy, 3 for an Scd galaxy, and 4 for an Im galaxy.
We then interpolate between the CCW spectral energy distributions
to construct the spectral energy distribution appropriate for
the given spectral type.  Galaxies whose colors are redder than
a CCW E galaxy or bluer than a CCW Im galaxy are simply assigned
the spectral energy distribution of an E galaxy or an Im galaxy,
respectively.  The number of galaxies with colors outside the limits
defined by the CCW E and Im types is known to be small even to
large redshifts (e.g., Crampton et al. 1995).  The fact that many
of our galaxies lie outside the CCW limits (see Figure 17, Paper I)
is due to the large errors in our colors ($\sigma_{g-r} \approx 0.35^m$).

We compute the absolute rest-frame $B_{AB}$-band magnitude as follows:
\begin{eqnarray}
M(B_{AB})_{\rm rest} & = & M(r)_{\rm rest} + (B_{AB}-r)_{\rm rest} \nonumber \\
		& = & r_{\rm obs} - 5\log_{10}D_L(z) - 25. - k_r(z) +
		      (B_{AB}-r)_{\rm rest} \\
		& = & r_{\rm obs} - 5\log_{10}D_L(z) - 25. + 
		      2.5\log_{10}(1+z) - k_{eff}, \nonumber
\end{eqnarray}
where $k_{eff}$ incorporates the corrections based on the spectral
energy distribution and $D_L$ is the luminosity distance in Mpc.
The $2.5\log_{10}(1+z)$ term represents the change in the bandwidth
with redshift and is included in the traditional $k$-correction.
Again following Lilly et al. (1995), we separate the bandwidth stretching
term, which has negligible error since it depends only on the accurately
measured redshift, from the terms which depend on the spectral
energy distribution and are therefore much more uncertain.  We plot
$k_{eff}$ for the $g$ and $r$ bands in Figure \ref{figures:keff}.
By converting from $g_{\rm obs}$ for objects with $z \lesssim 0.3$ and
from $r_{\rm obs}$ for objects with $z \gtrsim 0.3$, $k_{eff}$ may
be kept less than $\pm0.6^m$ for $z < 0.7$ for all spectral types.

\subsection{Method}

We use the step-wise maximum-likelihood (SWML) method of
Efstathiou, Ellis, \& Peterson (1988) to estimate the LF.  The probability
of observing a galaxy of absolute magnitude $M_i$ at redshift $z_i$
in a flux-limited catalog is given by,
\begin{equation}
p_i \propto {\phi(M_i) \over {\int_{-\infty}^{M_{max}(z_i)} \phi(M)dM}},
\end{equation}
where $\phi$ is the LF and $M_{max}(z_i)$ is the
intrinsically faintest galaxy observable at $z_i$ in the flux-limited
catalog.  The LF is parameterized as a set of $N_p$ numbers
$\phi_k$ such that
\begin{equation}
\phi(M) = \phi_k, \quad M_k - \Delta M/2 < M < M_k + \Delta M/2,
	\quad k = \ 1, \dots, N_p,
\end{equation}
and then the likelihood,
\begin{equation}
{\cal L} = \prod_{i = 1}^{N} p_i,
\end{equation}
where $N$ is the number of galaxies in the sample, is maximized with
respect to the $\phi_k$.  We constrain the values of $\phi_k$ to satisfy
\begin{equation}
\sum_{k = 1}^{N_p} \phi_k \Delta M = 1.
\end{equation}
The virtues
of the SWML method are that it is not biased by the presence of
clustering since the normalization of the LF cancels
out of the expression for the probability $p_i$ and also that one does
not have to assume a particular functional form for the LF.
In order to include
the weights, we make the substitution
\begin{equation}
{\cal L} = \prod_{i = 1}^{N} p_i \rightarrow {\cal L} = \prod_{i = 1}^{N}
{p_i}^{w_i},
\end{equation}
where $w_i$ is the weight of galaxy $i$ (Zucca, Pozzetti, \& Zamorani 1994,
Lin et al. 1996b).
One must then estimate the mean galaxy density separately.  We use
a standard technique which we describe below.
Since the weights are greater
than one, their use will increase the calculated likelihood for
the sample and thus lead to artificially small error estimates.  By
renormalizing the weights so that $\sum_{i = 1}^{N} w_i = N$, the
error estimates are appropriate for the true sample size (Lin et al. 1996b).
Note that the normalization constraint on the $\phi_k$ reduces
the estimated errors.

We do not use the
traditional $1/V_{max}$ method (\cite{Schmidt:vmax}) employed by
Ellis et al. (1996) and Lilly et al. (1995) since the method is sensitive
to clustering.
We have, however, compared the results
of the two techniques for samples with $z > 0.2$, where the clustering
in our survey is not pronounced, and found that they agree satisfactorily.
For $z < 0.2$, we can construct volume-limited sub-samples with
$r \le 20.0^m$ in which any galaxy with $M(B_{AB}) \le -19^m + 5\log_{10}h$
is visible in the entire volume.  Of course, the value of the
LF in a given magnitude bin
for a volume-limited sample is estimated by counting the number galaxies
with absolute magnitudes in the bin and then dividing by the volume
of the sample and the width of the bin.  The SWML LFs
for $z < 0.2$ agree well with the LFs estimated from
the volume-limited samples.

We compute the mean density $\bar n$ of a magnitude-limited sub-sample using
the following estimator:
\begin{equation}
\bar n = {1 \over V} \sum^{N}_{i = 1} { w_i \over s(z_i)},
\end{equation}
where $V$ is the volume of the sample, $N$ is the number of objects
in the sample, and $s$ is the selection function.  The selection
function,
\begin{equation}
s(z_i) = {\int^{M_{lim}(z)}_{-\infty} \phi(M) dM \over
{\int^{M_{max}}_{-\infty} \phi(M) dM}},
\label{equations:mean_density}
\end{equation}
gives the fraction of the LF observable at a given redshift.
Here, $\phi(M)$ is the LF, $M_{lim}(z)$ is the maximum
absolute magnitude that an object can have at redshift $z$ and still
be included in the sample, and $M_{max}$ is the absolute magnitude
of the most instrinsically faint galaxy in the sample.  In practice,
one does not begin evaluating the integrals at $-\infty$, but rather
at the absolute magnitude of the most instrinsically bright galaxy
in the sample.  The estimator in Equation \ref{equations:mean_density}
is almost identical to the minimum variance estimator derived
by Davis \& Huchra (1982) for $s \gtrsim 0.1$
and is unbiased by density inhomogeneities.

An additional complication of
computing a LF in the $B_{AB}$ band where the objects
have been selected in the $r$ band is that one must ensure that
any object, regardless of its color, would have been detectable in both bands.
If one ignores this complication, then the faintest objects at
a given redshift will be biased in color.  In our survey, since
the $r$ band is centered at a longer wavelength than the $B$ band,
the faintest objects would be biased to the red.  In order to
avoid such a bias, we adjust our absolute $B_{AB}$ magnitude limits
as a function
of redshift so that the bluest galaxy at any $B_{AB}$ magnitude limit would
be observable in the $r$ band.  For the local LF
computed in the $r$ band, the bias works in the opposite sense, and
so we adjust our rest-frame absolute $r$ magnitude limits to ensure that the
reddest galaxy a given limit would be detected in the observed $r$ band.

For each LF, we estimate the parameters of the
best-fitting Schechter (1976) function,
\begin{equation}
\phi(M)dM = \phi^\ast e^{-e^{.92(M^\ast - M)}+.92(M^\ast -M)\alpha},
\label{equations:Schechter}
\end{equation}
where $\phi^\ast$ is the normalization, $M^\ast$ determines the location
of the bright-end exponential cutoff, and $\alpha$ is the faint-end slope.
The fitting was performed
using a standard $\chi^2$-minimization algorithm (Press et al. 1992) with the 
Schechter function
integrated over the width of the adopted magnitude bin.  We intend these
fits to be useful for comparisons with other work.  Usually, there are too few
points for the fits to be well defined.

Our random magnitude errors ($\sigma \approx 0.25^m$) will artificially
brighten the characteristic magnitude $M^\ast$ of the LF and steepen
the faint end.  We correct for the magnitude errors by fitting to 
the data points a
Schechter function (Equation \ref{equations:Schechter}) convolved
with a Gaussian of dispersion $0.25^m$ (Efstathiou et al. 1988).
In fact, however, the corrections to the Schechter function parameters
are substantially less than the 1$\sigma$ statistical errors.

The error in the normalization
from the $\chi^2$-fitting only includes the uncertainties due to
$M^\ast$ and $\alpha$.  The error due to large-scale structure
fluctuations is
\begin{equation}
{\delta\bar n \over \bar n} \sim {J_3 \overwithdelims() V}^{1/2},
\label{equations:delta_n}
\end{equation}
(Davis \& Huchra 1982), where $J_3$ is the second moment of
the two-point spatial correlation function (Peebles 1980) and $V$
is the appropriate volume.  We use $J_3 \approx 10,000$ ($h^{-1}$ Mpc)$^3$
(Tucker et al. 1997) and record the error from density fluctuations
alongside the uncertainty in $\phi^\ast$ due to $M^\ast$
and $\alpha$.

\section{Results}

In the following subsections, we report our results for the local
LF, the evolution of the LF, and
the LFs of the Corona Borealis and Abell 2069 superclusters.
All samples are magnitude-limited at $r = 20^m$.  
The parameters of the best fitting Schechter functions are summarized
in Table 1, where the sample is given in the first column,
the number of galaxies used to compute the LF in the second column,
the absolute range over which the fit is valid in the third column,
$\phi^\ast$ in the fourth column, $M^\ast$ in the fifth column,
$\alpha$ in the sixth column, the reduced $\chi^2$ in the seventh
column, and the estimate of the variance due to density fluctuations
in the eighth column.  The numbers of galaxies listed in the second
column are slightly smaller than the total number of galaxies satisfying
the sample listed in the first column
because a few galaxies have been trimmed from each sample, as described
above, to ensure that there are no color biases in the faintest bins.
Corrections to $\phi^\ast$ to match galaxy counts,
as discussed below, have {\it not} been applied to the values
listed in Table 1.
We wish to emphasize that the
fitted Schechter functions are intended only to guide the eye and that
comparisons of the various LFs in this paper are best
done by comparing the individual data points in the plots.

\subsection{The Local Luminosity Function}

The $B_{AB}$-band local galaxy LF is plotted
in Figure \ref{figures:local_lf}.  The unfilled circles show the
LF for $z \le 0.2$ with the superclusters removed.
The filled circles show the LF for $z \le 0.2$ with the
the superclusters included.  In order to remove the superclusters,
we simply delete all objects with $0.06 \le z \le 0.13$.  The median
redshift of our local sample with the superclusters removed is $z_{med} = 0.15$.
The Stromlo/APM LF is plotted with the solid line.
We also plot the Autofid local LF with the dashed line. 
In Figure \ref{figures:local_lf} and all subsequent figures where
appropriate,
we convolve the Schechter function fits to the LFs from
other surveys with a Gaussian of dispersion $0.25^m$ in order
to facilitate comparisons with our LF data points, which are
constructed with galaxies whose photometry suffers from random
magnitude errors of $0.25^m$.
All of the local luminosity
functions have similar shapes, but the normalizations and low luminosity
ends vary significantly.

In order to investigate further the normalization of the local luminosity
function, we compute the {\it shape} of local luminosity in the rest-frame
$r$ band and then normalize this LF to the $r$-band
counts of Weir, Djorgovski, \& Fayyad (1995).  We plot our $r$-band local
LF, normalized to the counts of Weir et al. (1995), in Figure 
\ref{figures:local_lf_r}, along
with the $r$-band local LF from LCRS.  To convert from isophotal
$R_{LCRS}$ magnitudes to total Gunn $r$ magnitudes, we apply
a 25\% isophotal-to-total light correction and then use $R_{LCRS,total}
- r \approx 0.25^m$  (Shectman et al. 1996).   Thus, the corrections compensate
for each other and $R_{LCRS} \approx r$.
The counts of Weir et al. (1995) are based
on 4 overlapping, high galactic latitude plates taken as part of the
POSS-II survey.  Knowing the shape of the LF, we
can estimate the differential number counts as
\begin{eqnarray}
{dn \over dm} & = & \int_0^\infty \phi[M(m,z)] {dV \over dz} dz \nonumber \\
              & = & \phi^\ast \int_0^\infty \phi^\prime[M(m,z)] {dV \over dz} dz \\
	      & = & \phi^\ast {dI \over dm} \nonumber,
\label{equations:diff_counts}
\end{eqnarray}
where $\phi^\prime$ is $\phi$ with $\phi^\ast$ set equal to 1,
$M(m,z)$ is the absolute magnitude of an object at redshift $z$ with
apparent magnitude $m$, and the $m_i$ are the magnitude intervals.
We estimate $\phi^\ast$ by minimizing the quantity
\begin{equation}
\sum_i {{[dn(m_i) - \phi^\ast dI(m_i)]^2} \over {\phi^\ast dI(m_i)}}
\end{equation}
with respect to $\phi^\ast$ (Efstathiou et al. 1988), which yields
\begin{equation}
(\phi^\ast)^2 = {{\sum_i [dn(m_i)]^2 / dI(m_i)} \over {\sum_i dI(m_i)}}.
\end{equation}
We find $\phi^\ast = 1.5^{+0.7}_{-0.5} \times 10^{-2} h^3$ Mpc$^{-3}$ 
where the errors reflect the
changes due to varying jointly $M^\ast$ and $\alpha$ by $\pm1\sigma$.
This is a 21\% reduction from the normalization determined in the Norris
field itself.  Although the median redshift of our local sample
is $z_{med} = 0.15$, the agreement with LCRS ($z_{med} \approx 0.1$) in the
$r$ band leads us to believe that we have computed a fair estimate
of the local LF.

In Figure \ref{figures:r_counts}, we plot the $r$-band differential number
counts.  The histogram shows the counts from the POSS-II plate from which
we selected our objects.  The thick solid line is the counts from
Weir et al. (1995), and the triangles are CCD counts from Metcalfe et al. (1991).
The dashed line represents the predicted counts based
on our field galaxy LF, including evolution at
$z > 0.2$ (see \S 4.2 below), with the superclusters removed.
Even with the superclusters removed, our local LF appears
to be normalized too high.  The solid line, however, gives the predicted
counts with the normalization reduced by 21\%.  With the reduced normalization,
the predicted counts agree quite well with the observed counts to
$r = 20^m$.  For comparison, we also show, as the dotted line, the
predicted counts from the LCRS.  These counts fall below the Weir et al.
(1995) counts for $r \ge 18^m$ because they do not include the evolution
of the LF beyond $z \gtrsim 0.2$.
Since the volume of the Norris region, with the superclusters
removed, is $\approx 2.6 \times 10^5 h^{-3}$ Mpc$^3$, we would expect
from equation \ref{equations:delta_n} $\delta\bar n / \bar n \approx 0.20$.  
It is therefore not cause for concern that this region is overdense by 21\%.

\subsection{The Luminosity Function to $z = 0.5$}

We compute the field galaxy LF in two
redshift intervals: $0 < z < 0.2$ and $0.2 < z < 0.5$.  The
results are plotted in Figure \ref{figures:field_lf}.  The normalization
of the local LF (unfilled circles) has been reduced by 21\%
to match the $r$-band counts.  The filled circles are the
LF of the high redshift interval.  We also
plot the LFs of the Stromlo/APM survey (solid line),
the CNOC survey ($0.2 < z < 0.6$, dashed line), and the CFRS
survey ($0.2 < z < 0.5$, dotted line), all of which have been
convolved with a Gaussian with $\sigma_m = 0.25^m$ to account
for random photometry errors.
Since the LF for the
Autofib survey was divided into redshift intervals which do not
neatly match our redshift intervals and, more importantly, since
Lin et al. (1996a) have already performed a detailed comparison
with the Autofib survey, we do not plot the Autofib
LFs.  The $0.2 < z < 0.5$ LF
has clearly evolved with respect to the local LF.

\subsection{The Luminosity Function to $z = 0.5$ Divided by $W_0$([\ion{O}{2}])}

We create two sub-samples of galaxies according to the rest-frame equivalent
width of [\ion{O}{2}] $\lambda$3727.  The division is made at a rest-frame
equivalent width of 10\AA, which roughly corresponds to dividing the
sample into types earlier and later than Sbc (Kennicutt 1992) and
thus allows direct comparison with the results of CNOC and CFRS.
For galaxies with $z < 0.049$, [\ion{O}{2}] is not redshifted into
our observed wavelength range, and so we use the strength of H$\beta$
to divide our sample.  After correcting for stellar absorption,
we have 7 galaxies with $z < 0.049$ and $W_0$(H$\beta$) $>$ 5\AA,
which we include in the $W_0$([\ion{O}{2}]) $>$ 10\AA\ sample. 
We have also investigated separating our sample by color, but we have
found that the large errors on our colors tend to dilute trends
which are seen clearly in samples defined by $W_0$([\ion{O}{2}]).
The LFs for galaxies with $W_0$([\ion{O}{2}]) $<$ 10\AA\ and
$W_0$([\ion{O}{2}]) $>$ 10\AA\ are shown in Figure \ref{figures:no_o2_lf} and
Figure \ref{figures:o2_lf}, respectively.  We also plot
in each figure the LFs for the corresponding color-selected samples
from CNOC and CFRS.  It is important to remember when considering
possible detailed discrepancies between our LFs and the CNOC and CFRS LFs that
our sample is divided by $W_0$([\ion{O}{2}]), which, while roughly
equivalent to color selection, is not identical.
There is no significant indication that the population of early-type galaxies
(i.e.,
$W_0$([\ion{O}{2}]) $<$ 10\AA) has evolved since $z = 0.5$; this result is not
surprising given that the light of early-type galaxies is
dominated by red, long-lived stellar populations.  In contrast,
the LFs of the late-type galaxies (i.e., those with
$W_0$([\ion{O}{2}]) $>$ 10\AA) show striking evidence for evolution, even
though the sample sizes are small and the error bars are large.

\subsection{The Supercluster Luminosity Functions}

The LFs of the Corona Borealis supercluster and
the Abell 2069 supercluster are given in Figure \ref{figures:clusters_lf}.
We take the redshift range of the Corona Borealis supercluster to
be $0.06 \le z \le 0.09$ and that of the Abell 2069 supercluster to
be $0.10 \le z \le 0.13$.
The normalization of the Corona Borealis supercluster function is a
factor of 2 greater than that of the Abell 2069 supercluster, but the
shapes of the LFs of the two superclusters are similar.  Note that
the volumes used to normalize the superclusters LFs are in redshift
space; the real-space volumes may be quite different (see \S 5.3
below for a detailed discussion).  Since the
bright ends of the supercluster LFs are clearly not well-described
by a Schechter function, we restrict our fit to $M(B_{AB}) > -21.3^m$.
In addition, the fit to the Corona Borealis supercluster is limited
to $M(B_{AB}) < -17.1^m$ since the faintest two data points appear
to describe a sharp upturn in the LF.

\section{Discussion}

\subsection{The Local Luminosity Function}

Prior evidence for rapid evolution of the galaxy LF 
to $z \sim 0.1$ from galaxy counts
was based crucially on normalizing the local LF to the
bright ($B_{AB} \sim 16^m$) galaxy counts from Schmidt-telescope photographic
surveys (e.g., Maddox et al. 1990).  With this low normalization,
predicted counts from the no-evolution model fall well short of the
observed counts for $B_{AB} > 18^m$.  However, the evolution of the
LF which we and others observe is not enough
to make up for this shortfall.  In Figure \ref{figures:B_counts},
we plot the observed
counts from the APM survey (Maddox et al. 1990) and from the CCD survey
of Metcalfe et al. (1991), along with various predicted counts.
The dotted line shows the expected counts using the Loveday et al.
(1992) LF for $0 < z < 0.2$ and the CNOC $B_{AB}$
LF for $z > 0.2$.  Despite including the observed
evolution of the LF, the predicted counts only
begin to agree with the observed counts for $B_{AB} \gtrsim 19^m$,
by which point the predicted counts are dominated by galaxies with
$z > 0.2$.  In contrast, the predicted counts computed using
the evolving LFs measured for our survey and
for the Autofib survey, while substantially overpredicting the
counts for $B_{AB} \lesssim 18^m$, match the observed counts for
$18^m \lesssim B_{AB} \lesssim 20.5^m$.

In order to help unravel this confusing situation, we plot
in Figure \ref{figures:B_and_r_lf}
various $B_{AB}$- and $r$-band local LFs on the
same diagram.  $B_{AB}$-band LFs are plotted with
respect to the bottom axis, while $r$-band LFs are
plotted with respect to the top axis.  The two axes are offset
by $B_{AB} - r = 0.72^m$, which is the median color we measure
for field galaxies with $z < 0.2$.  Our $B_{AB}$- and $r$-band
local LFs, both of which have been reduced by 21\%
following the discussion in \S 4.1, are plotted with filled and unfilled
circles, respectively.  With the color offset, they agree extremely
well.  The Stromlo/APM LF, represented by the solid line, lies
consistently below our $B_{AB}$-band LF.  The LCRS $r$-band LF
is consistent with our $r$-band LF.  Unlike Lin et al. (1996b), we
do {\it not} conclude that the LCRS $r$-band LF matches the
Stromlo/APM LF.  The reason for the disagreement lies in the
different measurements of the median color of local galaxies.  Our
median color is that of an Sb galaxy, whereas the LCRS median color
is that of a much redder E galaxy.  The median color of galaxies
in the {\it Third Reference Catalogue of Bright Galaxies}
(de Vaucouleurs et al. 1991) is $\langle B - V \rangle \approx 0.75^m$
(see Table 2 of Fukugita et al. 1995), which is roughly that of
an Sb galaxy.  Sebok (1986) also concludes that the typical local
galaxy has the color of an Sb galaxy.  It is thus surprising that
the mean color of the LCRS galaxies is so red.  Since our $r$-band
LF agrees with the LCRS $r$-band LF and since the LCRS computed the
colors of their galaxies by matching directly to the APM catalog,
the most natural explanation
for the anomalous red colors of the LCRS is a systematic error in the
bright APM magnitudes.  We note that Weir et al. (1995) conclude that
magnitudes derived from $J$ plates are only reliable for $g > 16^m$
or, equivalently, $B_{AB} > 16.5^m$.  Galaxies brighter than $B_{AB} \approx
16.5^m$ are saturated on the photographic plates.
A systematic error in the
bright APM counts would remove the need for rapid, and otherwise
unsubstantiated, galaxy evolution at $z \sim 0.1$.

We note also that there is possible evidence for a rise in the
local LF above an $\alpha \approx -1$ for the least luminous galaxies
in our survey.  Such a rise is evident in the data of Marzke et al. (1994b)
for irregular galaxies,
the CFRS, and the low surface brightness galaxy redshift survey of
Sprayberry et al. (1997).  Similar behavior is also perhaps visible
in the local LF of the Autofib survey.  Although Ellis et al. (1996)
argue against a rise in the LF for $M(B_{AB}) > -16^m$, the three
faintest points in the their local LF (their Figure 8) all lie
above their preferred Schechter function fit.
However, since such faint galaxies are only visible
in our survey in a quite small volume, we refrain from attempting to
make a definitive statement.

\subsection{The Evolution of the Luminosity Function to $z = 0.5$}

We asserted in \S 4.2 that the LF of star-forming
galaxies ($W_0$([\ion{O}{2}]) $>$ 10\AA)
evolved from $z = 0$ to $z = 0.5$ and that the luminosity
function of galaxies with weak [\ion{O}{2}] emission ($<$ 10\AA) did not.  A powerful method to verify this result
is to compute $\langle V/V_{max} \rangle$ for appropriate samples
(\cite{Schmidt:vmax}).  If there is no evolution in the number density
of objects, $\langle V/V_{max} \rangle = 0.5$; if the number density
declines, $\langle V/V_{max} \rangle < 0.5$; and if the number density
increases, $\langle V/V_{max} \rangle > 0.5$.  Our $\langle V/V_{max} \rangle$
analysis is complicated by the need to remove the superclusters and to
account for the 21\% overdensity of our local field.  In order to excise
the superclusters, we simply remove all galaxies with $0.06 \le z \le 0.13$.
If the maximum redshift $z_{max}$ at which a galaxy could be observed
in our survey lies in the range 0.06 to 0.13, we set $z_{max} = 0.06$.
We correct for the overdensity of our local field by reducing the weights
of galaxies with $z \le 0.2$ by 21\%.
The values of $\langle V/V_{max} \rangle$ for various samples of
galaxies are given Table 2, both with and without
the 21\% correction to the weights of the galaxies with $z \le 0.2$.
For samples selected by color, we compute the $\langle V/V_{max} \rangle$
statistic for galaxies in the redshift range $0 < z \le 0.5$ (with
the supercluster region excluded).  For samples selected by the
strength of [\ion{O}{2}], we use the redshift range $0.049 < z \le 0.5$
(with the supercluster region excluded) since [\ion{O}{2}] from objects
with $z < 0.049$ is not redshifted into our observed wavelength range.
The differences between our weighted and unweighted statistics are small.
The $\langle V/V_{max} \rangle$ test supports our
claim, at the $3.5\sigma$ level for the weighted statistic, that the population
of star-forming galaxies is evolving.
The rate of evolution increases with the strength of [\ion{O}{2}].  The
population of galaxies
with $W_0$([\ion{O}{2}]) $>$ 20\AA\ has $\langle V/V_{max} \rangle = 0.65
\pm 0.04$.  This result is analogous to the results from CFRS in which
the rate of evolution is the strongest for the bluest population of galaxies.
$\langle V/V_{max} \rangle$ for the population of red galaxies is consistent with
no evolution, in agreement with the LF analysis.

Broadly speaking, our results are in accord with the results of
CFRS, Autofib, Cowie et al (1996), and CNOC.  Since the CNOC survey,
like our survey, uses photometry in the $g$ and $r$ bands, it is particularly
interesting to compare our results in detail to theirs since many of
the systematic effects associated with $k$- and color-corrections ought
to be the same.  It is encouraging to see (Figure \ref{figures:no_o2_lf} and
\ref{figures:o2_lf}) that, given
the small samples, our LFs agree well with
those of CNOC.  The agreement is significantly improved if one reduces
the normalization of the CNOC LFs by 20\%, as, in fact, is recommended
by Lin et al. (1996a).

Now that we have confirmed that the population of blue galaxies is evolving
with redshift, we wish to investigate whether we can detect differences
in the colors and the spectral properties of the evolving population with
redshift.
First, we reiterate that the color distribution of objects with measured
redshifts is similar to the color distribution of unidentified objects, leading
us to believe that the type distribution of the identified objects
is not strongly biased (Paper I).  Although emission lines are generally easier to detect
than absorption lines, the difficulty of identifying emission lines at
observed wavelengths longer than 5577\AA, where there are many strong
night sky features, combined with the strength of Ca H, Ca K, and the 4000\AA\ break
in absorption line objects at $0.4 \lesssim z \lesssim 0.6$
mitigate the bias in favor of emission line objects. A sample
of the spectra of 8 absorption line objects in this redshift range is shown in
Figure \ref{figures:ab_line_objs} to illustrate the ease of detection of their
characteristic absorption features.

In Figure \ref{figures:obs_g-r}, we plot the observed $g-r$ color as
a function of redshift
of all the objects in our survey along with the tracks of five
representative model galaxy spectra.  The bluest spectrum is simply a
flat-spectrum object, $f_\nu = 0.$  The four other spectra are typical
of the Hubble types E, Sbc, Scd, and Im and are taken from
CCW.  The large, solid diamonds mark the observed median
color in the redshift ranges $0.13 < z < 0.2$ (arranged to exclude the
superclusters), $0.2 < z < 0.3$,
$0.3 < z < 0.4$, $0.4 < z < 0.5$, and $0.5 < z < 0.6$.  Perhaps
counter-intuitively, we
see that the observed median color does not become progressively bluer with respect
to the model spectra with increasing redshift.  As discussed by Lilly
et al. (1995) and illustrated in our Figure \ref{figures:rest-frame_g-r},
the median color does not become bluer because the color-magnitude
relation in the local universe (i.e., the fact that more luminous galaxies
are redder) breaks down for $z \gtrsim 0.2$ as the population of blue galaxies
brightens while the population of red galaxies does not evolve significantly.
We plot in
Figure \ref{figures:rest-frame_g-r} the absolute $B_{AB}$-band magnitudes
versus rest-frame $g-r$ for our galaxies divided into four intervals
in redshift.  In the low-redshift interval, the color-magnitude
relation is apparent, but it disappears in the higher redshift intervals.
At $z \lesssim 0.2$, we observe far down the LF to
absolute magnitudes where blue galaxies are dominant.  At $z \gtrsim 0.2$,
we do not observe as far down the LF, but the population of blue
galaxies has brightened so as to be included in the samples.

The increase in the luminosity of the population of blue galaxies is 
presumably associated
with a change in the star formation activity at earlier times.  
In our spectra, there are
two convenient star formation indicators, [\ion{O}{2}] $\lambda$3727
and H$\delta$ $\lambda$4101.  
[\ion{O}{2}] emission is found in galaxies with ongoing
star formation, and its strength is proportional to the strength
of H$\alpha$ (\cite{Kennicutt:atlas}).  Strong H$\delta$ absorption is
a signature of the presence of a population of A-stars, which are visible
$\sim$1 Gyr after a burst of star formation.
These lines are reliably
measured by automated programs (see Paper I) since both occur in regions
of the spectrum where the continuum is featureless and there is little
crowding from other lines.   An important virtue of the
H$\delta$ line is that, since it appears in absorption, a galaxy with
detectable H$\delta$ would have been identified no matter what its
spectral characteristics, which implies that
there is no bias towards detecting objects with
H$\delta$ absorption.  A galaxy with [\ion{O}{2}] emission is, of course, 
easier to identify than if it had had only absorption lines.  However, as we discussed
above, the combination of the difficulty of identifying weak emission
lines in the face of strong sky subtraction residuals and of the ease
of identifying
the strong features characteristic of absorption-line galaxies
at moderate redshifts suggests that our survey is not strongly
biased towards detecting objects with [\ion{O}{2}].
By adding together
spectra of late-type galaxies in two redshift intervals in order to
obtain two spectra with very high signal-to-noise ratios, Heyl et al. (1996)
find that the two star-formation indicators [\ion{O}{2}] $\lambda$3727
and H$\delta$ have both become stronger in the higher redshift mean
spectrum.  Those authors interpreted the increase in the strength
of [\ion{O}{2}] and H$\delta$ in the $z > 0.2$ spectra to imply that
not only were the higher-redshift galaxies forming stars more rapidly,
but that also the nature of the star formation was changing with redshift.
Specifically, they asserted that the strong H$\delta$ absorption in the
higher redshift sample is evidence that the star formation in that
sample is dominated by bursts.
With our high-quality spectra,
we can reliably measure [\ion{O}{2}] and H$\delta$ without adding
together the spectra of many galaxies.
In Figure \ref{figures:o2_and_hd}, we plot as a function of redshift
the fraction of galaxies with
[\ion{O}{2}] emission and the fraction of galaxies with H$\delta$
absorption.  The vertical dashed line
marks the redshift beyond which H$\delta$
is shifted into the region of the spectrum in which sky subtraction
becomes increasingly difficult.  While the fraction of galaxies
with [\ion{O}{2}] emission increases with redshift, the
fraction of galaxies with strong H$\delta$ absorption shows no
significant variation with redshift, suggesting that the star formation
is {\it not} occurring in short-lived bursts (c.f., Hammer et al. 1997)

In order to investigate this disagreement more carefully, we
repeat the analysis of Heyl et al. (1996) and construct high
signal-to-noise ratio composite spectra.  The individual spectra
have been median-combined after scaling by the median count level.
The magnitude weights, which correct for our incompleteness at
faint magnitudes, have not been applied since scaling by the
median count level in each spectrum effectively incorporates the
magnitude weights.  In Figure \ref{figures:coadd}, we plot the composite spectra
of galaxies with $W_0$([\ion{O}{2}]) $>$ 20\AA, $M(B_{AB}) \le -19^m$,
and either $0.13 < z \le 0.2$ (thick line, $z_{med} = 0.16$)
or $0.2 < z \le 0.5$ (thin line, $z_{med} = 0.38$).  
$W_0$([\ion{O}{2}]) $>$ 20 \AA\ is typical for the
late-type galaxies in which Heyl et al. (1996) observe
spectral changes with redshift.
While the [\ion{O}{2}] line is modestly stronger in the
higher redshift composite spectrum, there is no difference in the
strength of H$\delta$ (which is partially filled-in by emission in both spectra), which indicates that the nature of the
star formation has not changed from $z = 0.5$ to $z = 0.$  In contrast
to Heyl et al. (1996), we again do not conclude that the spectra of
intermediate-redshift blue galaxies show spectral signs of short
bursts of star formation.  (We do not believe that the small differences
in the rest-equivalent widths of the Ca H and K lines are significant
since it is difficult to fit a reliable continuum in the vicinity
of the 4000\AA\ break.)

The modest increase in the strength of the [\ion{O}{2}] line with
redshift for galaxies with $M_{B_{AB}} \le -19^m$ is due, as previously
discussed by Cowie et al. (1996), to the increase in the intrinsic
luminosity of star-forming galaxies with redshift.
In Figure \ref{figures:o2_ew}, we plot
absolute magnitude versus the rest equivalent width of [\ion{O}{2}] for the
galaxies in our sample divided into four redshift intervals.  As
the redshift increases, more and more luminous galaxies exhibit
strong [\ion{O}{2}].  Notice, however, that the range of rest equivalent
widths does not vary significantly with redshift.

It is difficult to draw firm conclusions about the nature of the
evolving galaxy population from the LFs alone.
The LFs provide only a statistical view of the entire population
and include information about the evolution of individual galaxies
only indirectly.  Furthermore, the luminosity of a galaxy is an
unreliable tracer of the physical state of the galaxy.  The luminosity
of a galaxy, especially in the rest-frame ultraviolet, can change
dramatically on short timescales, making the identification of
the descendants of distant galaxies in the local population very
difficult.  In order to make further progress, additional data
on the nature of distant galaxies is required.  The morphologies
of distant galaxies, measured with the {\it Hubble Space Telescope} (HST),
are already providing crucial clues
(\cite{Driver:HST_counts}, \cite{Glazebrook:HST_counts},
\cite{Abraham:HST_counts}).  The counts of faint elliptical and
early-type spiral galaxies match predictions based on counts
in the local universe,
provided that the local LF is normalized, as
advocated here, a
factor of $\sim$1.5-2 higher than found by Loveday et al. (1992).  In contrast,
the HST number counts of late-type and irregular galaxies are far in
excess of the counts expected from observations of nearby galaxies, even with
a high normalization of the local LF.  

An alternative method for selecting distant galaxies is by their
gas absorption cross section.  Steidel, Dickinson, \& Persson (1994)
study the evolution of 58 galaxies selected by
\ion{Mg}{2} $\lambda\lambda$2796, 2803 absorption
seen in the spectra of high-redshift quasars.  
These authors found that the galaxies
responsible for quasar absorption lines in the range $0.2 \le z \le 1.0$,
which typically have luminosities near $L^\ast$, do not evolve over the
redshift range.  However, their sample is small and, when divided
by color, is not inconsistent with results from surveys of galaxies
selected by apparent magnitude (Lilly et al. 1995).  In addition,
Steidel et al. (1994) note that intrinsically faint blue galaxies
do not appear in their sample, and so the population of galaxies which is
observed to be evolving most rapidly is not included in their survey.

It is now possible, especially with 10m-class telescopes, to measure
the masses (e.g., Vogt et al. 1993, 1996; Rix et al. 1996, Guzm\'an et al. 1996)
and chemical abundances of distant galaxies.  Since the masses and
metallicities of galaxies evolve more smoothly than the luminosities,
measurements of these two quantities will allow more easily interpretable comparisons
of distant and local populations (Guzm\'an et al. 1996).  In addition,
the mass and metallicity are, compared to morphology and even gas
absorption cross section (Churchill, Steidel, \& Vogt 1996), straightforward
to define and interpret.  A survey to measure
the masses and chemical abundances of the faint
blue galaxies ought to yield important insights into the nature of
this rapidly evolution galaxy population and aid in the identification
of their present-day counterparts.

\subsection{The Supercluster Luminosity Functions}

The LFs of the two superclusters have quite similar
shapes and generally resemble the field galaxy LF.
Since
the normalizations of the superclusters LFs are computed in redshift
space, in which distances along the line-of-sight may be substantially
altered with respect to real space, 
it is not straighforward to compare the normalization of
the supercluster LFs with that of the local field galaxy LF.  For
the purpose of the discussion here, we introduce a factor
$f$, the ratio of the redshift-space volume to the real-space
volume.  We expect $f$ to be in the range $1 \lesssim f \lesssim 5$.
The lower limit corresponds to assuming that the peculiar velocities
in the supercluster regions are small; the upper limit corresponds
to assuming that the depth of the superclusters along the line-of-sight
is similar to the linear sizes of the superclusters on the plane of the
sky ($\sim 20h^{-1}$ Mpc).  The real-space mean density of galaxies in the
Corona Borealis supercluster, obtained simply by integrating the
measured LF data points ($-22.1^m \le M(B_{AB}) \le -16.3^m$), is
$\bar n_{CB} \approx 0.4f_{CB}^{-1} h^3$ Mpc$^{-3}$.  The mean density
of field galaxies in the same range of absolute magnitude is
$\bar n_{field} \approx 0.05 h^3$ Mpc$^{-3}$.  Similarly, the
real-space mean density of galaxies in the A2069 supercluster
($-21.7^m \le M(B_{AB}) \le -17.5^m$) is
$\bar n_{A2069} \approx 0.1f_{A2069}^{-1} h^3$ Mpc$^{-3}$, while
the mean density of field galaxies in the same range of absolute
magnitude is $\bar n_{field} \approx 0.04 h^3$ Mpc$^{-3}$.  Thus,
the overdensities are $\sim 40f_{CB}^{-1}$
and $\sim 13f_{A2069}^{-1}$ for the Corona Borealis and A2069 superclusters,
respectively.

It is also important to know whether our sampling of the superclusters
is biased towards either the Abell clusters within the superclusters
or the ``field'' of the superclusters.  Owing to the difficulties
of converting redshift-space volumes to real-space volumes, we
assess our sampling of the superclusters using projected surface densities,
which are not affected by redshift-space distortions.  For the Corona
Borealis supercluster, the project surface density $\Sigma_{CB}$
is the mean redshift-space galaxy volume density
multiplied by the depth in redshift space of the supercluster along the
line-of-sight,
$c\Delta z/H_0 = 90 h^{-1}$ Mpc.  Thus, $\Sigma_{CB} \approx 0.4 \times 90
h^2$ Mpc$^{-2}$ $= 36 h^2 $ Mpc$^{-2}$.  We compare this value
with the median surface density of the regions surrounding successfully
observed Corona Borealis supercluster galaxies.
We compute the surface density around a given Corona Borealis supercluster galaxy
by counting the number of supercluster
galaxies in a 6 arcmin diameter circle surrounding the chosen galaxy.
We either simply count the number of galaxies with measured redshifts
within the Corona Borealis supercluster, or we count the total number
of galaxies on the original POSS-II $F$ plate, weighted by the
empirically-determined
fraction of galaxies at a given magnitude which are in the supercluster.
The results of these two methods
agree well:  the median values of the raw and weighted surface densities
are $\Sigma_{raw,med} = 27 h^2$ Mpc$^{-2}$ and
$\Sigma_{weighted,med} = 42 h^2$ Mpc$^{-2}$.  These two values bracket
the projected surface density measured by multiplying the mean
galaxy density (computed from the LF) by the line-of-sight depth
of the supercluster.  We conclude, therefore, that we have fairly sampled
the Corona Borealis supercluster.

We perform an identical analysis for the background
A2069 supercluster.  The redshift-space depth of the A2069
supercluster is also $90 h^{-1}$ Mpc.
Given a mean galaxy
volume density of $\bar n_{A2069} \approx 0.1 h^3$ Mpc$^{-3}$,
the projected surface density is $\Sigma_{A2069} = 9 h^2$ Mpc$^{-2}$.
The median surface density of the regions surrounding the successfully
observed A2069 supercluster galaxies, computed in the same fashion
as for the Corona Borealis supercluster, is $\Sigma_{med,raw} = 13
h^2$ Mpc$^{-2}$ (unweighted) or $\Sigma_{med,weighted} = 16 h^2$
Mpc$^{-2}$ (weighted).  Thus, we are slightly biased to the denser
regions of the A2069 supercluster.

The overall resemblance between the supercluster LFs
and the field galaxy LF suggests that 
the fundamental physical processes which drive galaxy formation
and evolution must not depend strongly on environment.  There are,
however, important differences between the LFs
in the field and in the superclusters that must ultimately be due
to environmental effects.
The most striking difference between the field and supercluster
LFs is that the supercluster LFs to do not continue the exponential
decline for galaxies brighter than $M(B_{AB}) \lesssim -21^m$.
Both superclusters evidently contain a population of very luminous
galaxies.  Despite the fact that there are only 6 galaxies with
$M(B_{AB}) < -21.3^m$ in the two superclusters combined, it is clear that
these galaxies are giant ellipticals found in the densest regions
of the superclusters.  The 4 of the 6 for which we have spectra (the
other 2 were taken from the literature) are dominated by the light
of an old, red stellar population.  All of the galaxies are found
in regions in the upper 27th percentile of local surace density, with half
of them found in regions in the upper 10th percentile.  In fact,
the four galaxies in the Corona Borealis supercluster are all found
in the dense ridge of galaxies between the Abell clusters A2061 and
A2067.
For the Corona Borealis supercluster, the characteristic magnitude
$M^\ast$ is $\sim 0.5^m$ brighter than in the field and is quite
close to the value measured by Colless (1989) for rich clusters.
There is
no significant difference between $M^\ast$ for the Abell 2069
supercluster and that of the field.   

For the Corona Borealis supercluster LF, the data points for the two faintest
magnitude bins hint that the LF may steepen significantly for galaxies fainter
than $M(B_{AB}) \gtrsim -17^m$.  Since the hint is based on only two
data points, which are themselves based on only 29 galaxies, we must be cautious
in our interpretation.  However, a steepening of the supercluster LF fainter
than $M(B_{AB}) \gtrsim -17^m$ would be in accord with observations of
the faint end of the LF in galaxy clusters and groups, in which a number of
workers report steep ($\alpha \lesssim -1.3$) LFs
(Impey, Bothun \& Malin 1988; Ferguson \& Sandage 1991; Biviano et al. 1995;
De Propris et al. 1995; Driver \& Phillipps 1996).
With the exception
of the study of the Coma cluster by Biviano et al. (1995), the observations
of steep faint ends in cluster and group LFs have depended, since redshifts
were not available, on the subtraction of a background component from
the foreground cluster, a procedure which is prone to systematic errors.
Although it would be unwise to draw strong conclusions from our two data
points, they do have the virtue of being based on galaxies with measured
redshifts.

\section{Summary}

We have presented an analysis of the LF of galaxies
in the Norris Survey of the Corona Borealis Supercluster.  Our
$r$-band LF of local field galaxies, when normalized to counts in
high galactic latitude fields, agrees well with the LCRS.  However,
the normalization of our $B_{AB}$ local LF is roughly a factor
of 1.6 higher than that of the Stromlo/APM survey.  Since
Lin et al. (1996b) claim that the LCRS local LF agrees well with
the Stromlo/APM survey, the difference must lie in a systematic
photometry error in one (or more) of the three surveys.  A clue
to the nature of this error is provided by examining the mean
colors of Norris and LCRS galaxies.  The mean color of local Norris
galaxies is that of an Sb galaxy, whereas the mean color of
LCRS galaxies, computed by matching LCRS galaxies directly with
galaxies in the APM catalog, is that of an E galaxy.  Given the
agreement of the Norris and LCRS $r$-band LFs, we therefore believe
that the error is most likely in the APM catalog.  Indeed, brightening
the magnitudes of APM galaxies with $b_j < 17^m$ by $\sim$0.25$^m$
would bring all of the local LFs into agreement.
A CCD-based local redshift survey (e.g.,
the Sloan survey, Gunn \& Weinberg 1995) will certainly resolve any remaining questions about the local LF.

We have observed evolution of the field galaxy LF within our sample,
thereby confirming the conclusions drawn from several previous
redshift surveys.     
The evolution is limited to the population of blue, star-forming galaxies.
The population of blue galaxies becomes more luminous with increasing redshift,
and thus the median color of the field galaxy population does not change.
The evolution of the population of blue galaxies is reflected in the larger
fraction of galaxies at higher redshift exhibiting spectral signatures
of ongoing star formation.  In contrast to the results of Heyl et al.
(1996), but in agreement with Hammer et al. (1996), we find that the
star formation is long-lived.  We do not see evidence for short-term
bursts of star formation.
We are unable to detect any evolution of the population of galaxies with 
$W_0$([\ion{O}{2}]) $<$ 10\AA.
The fact that the evolution which we observe in our $g$- and $r$-band
selected survey is consistent with the results of the surveys of Lilly et al. (1995), Ellis et al. (1996), Cowie et al. (1996), and Lin et al. (1996)
adds to the already strong evidence that a consistent picture
of the evolution of the galaxy LF is emerging.
In particular, it is quite reassuring that our LFs
agree well with those of Lin et al. (1996a) since both used the
$g$ and $r$ bands and should therefore have very similar systematic
effects.

The LFs of the two superclusters show significant
differences from the field galaxy LF, despite considerable
overall similarity.  Since the superclusters are $\sim 10 - 40 \times$
denser than the field, we are likely to be observing the influence
of the environment on galaxy formation and evolution.
The most prominent difference is an excess
of very bright galaxies ($M(B_{AB}) \lesssim -21^m$) relative to
the best-fitting Schechter function, which accurately describes
the field LF over the observed absolute magnitude range.  These very
bright galaxies are found in very dense regions of the superclusters
and have spectra dominated by an old, red stellar population.
In the
Corona Borealis supercluster, the characteristic magnitude $M^\ast$
is $\sim 0.5^m$ brighter than in the field.  $M^\ast$ for the
Abell 2069 supercluster is, however, very close to the value in
the field.  We have also presented suggestive evidence that
there is a sharp upturn in the supercluster LF for $M(B_{AB}) \gtrsim
-17^m$.  While there is also a suggestion of an upturn in the local field
galaxy LF for the least luminous galaxies in our survey, it does not
appear as dramatic as the upturn seen in the supercluster LF, but more
data are needed before the possible difference can be quantified.

\acknowledgements 

We are grateful to the Kenneth T. and Eileen L. Norris Foundation
for their generous grant for construction
of the Norris Spectrograph. 
We thank the staff of the Palomar Observatory for the expert assistance
we have received during the course of the survey, David Hogg for
many enlightening discussions, and the referee for a careful reading
of this paper and helpful suggestions.
This work has been supported by an NSF Graduate Fellowship (TAS) and
NSF grant AST-92213165 (WLWS).

\newpage
\clearpage

\begin{figure}
\plotfiddle{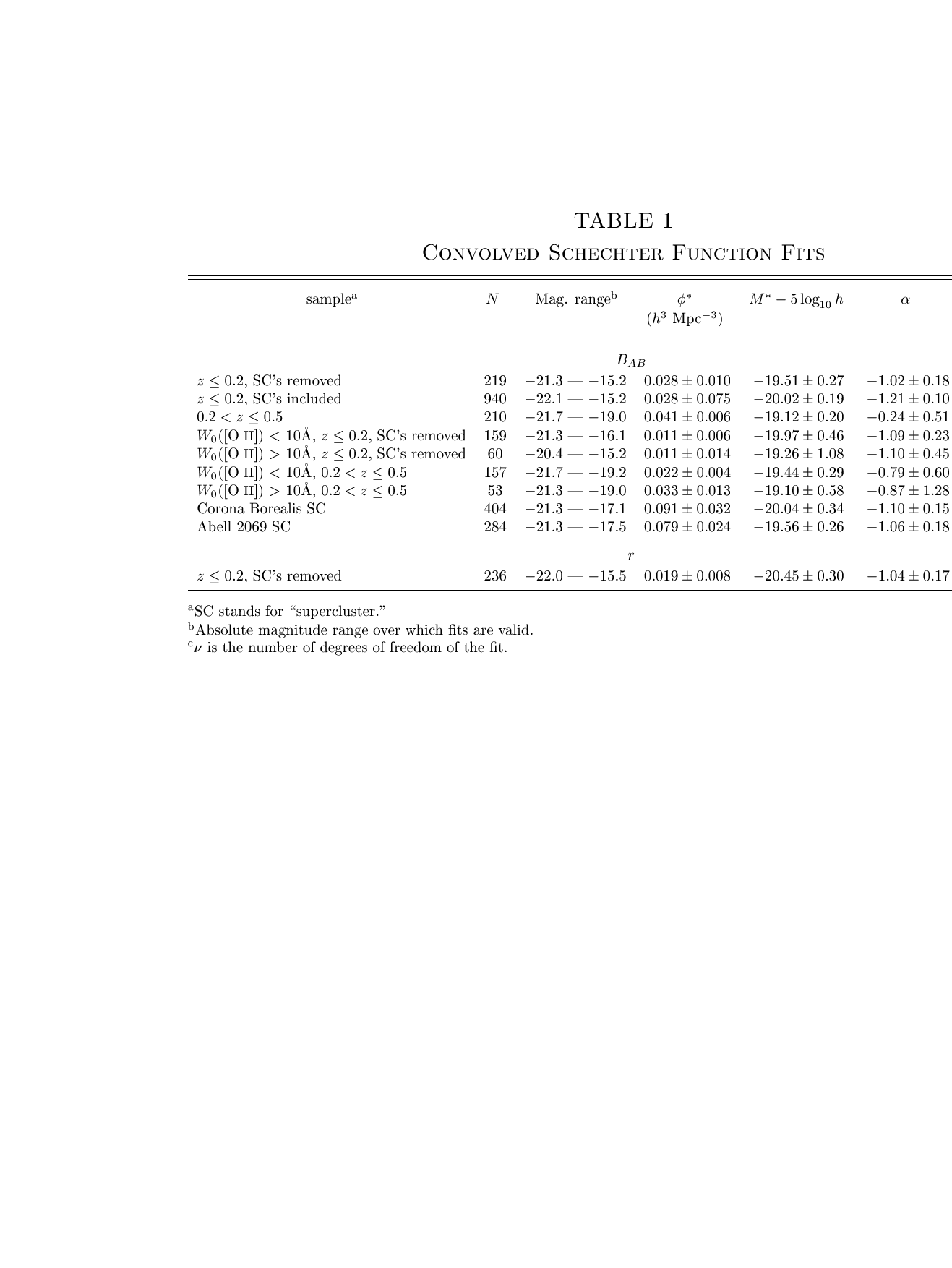}{6in}{90}{100}{100}{540}{-216}
\label{table:Schechter_fits}
\end{figure}

\newpage
\clearpage

\begin{figure}
\plotone{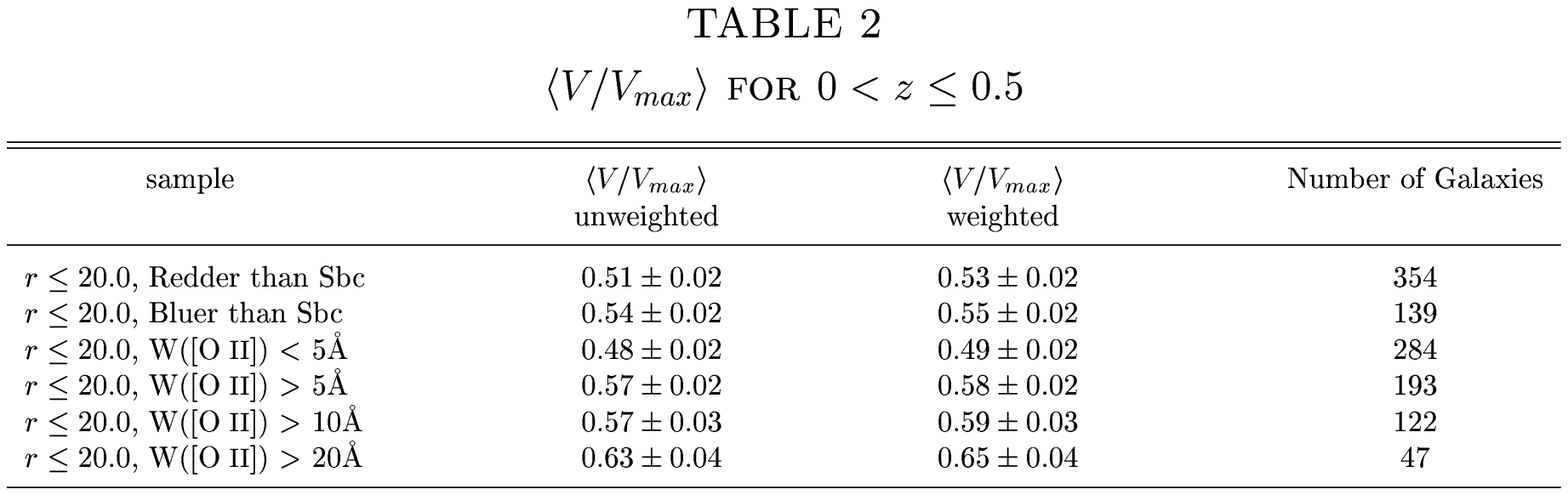}
\label{table:vmax}
\end{figure}

\newpage

\begin{figure}
\plotfiddle{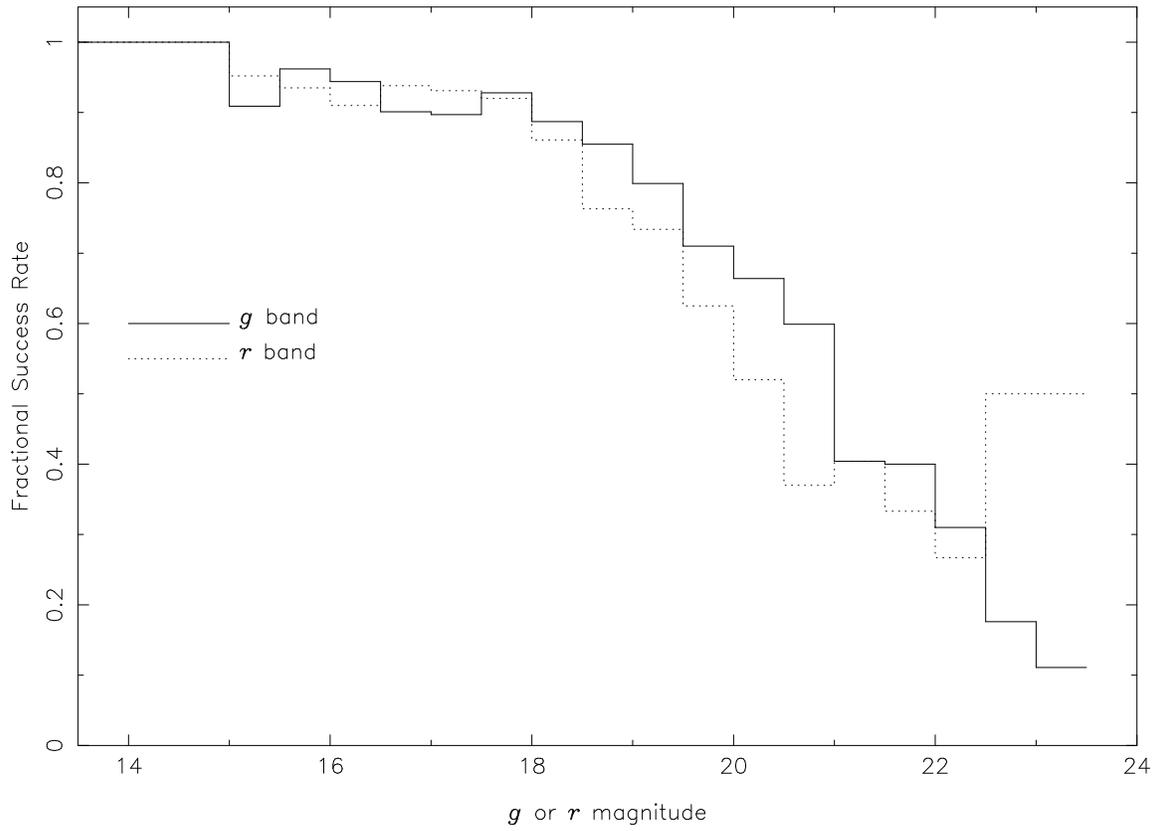}{4.3 in}{-90}{62}{62}{-234}{360}
\caption[]
{Redshift identification success rate as a function of $g$ (solid line)
and $r$ (dotted line) magnitudes.}
\label{figures:success}
\end{figure}

\newpage
\clearpage

\begin{figure}
\plotfiddle{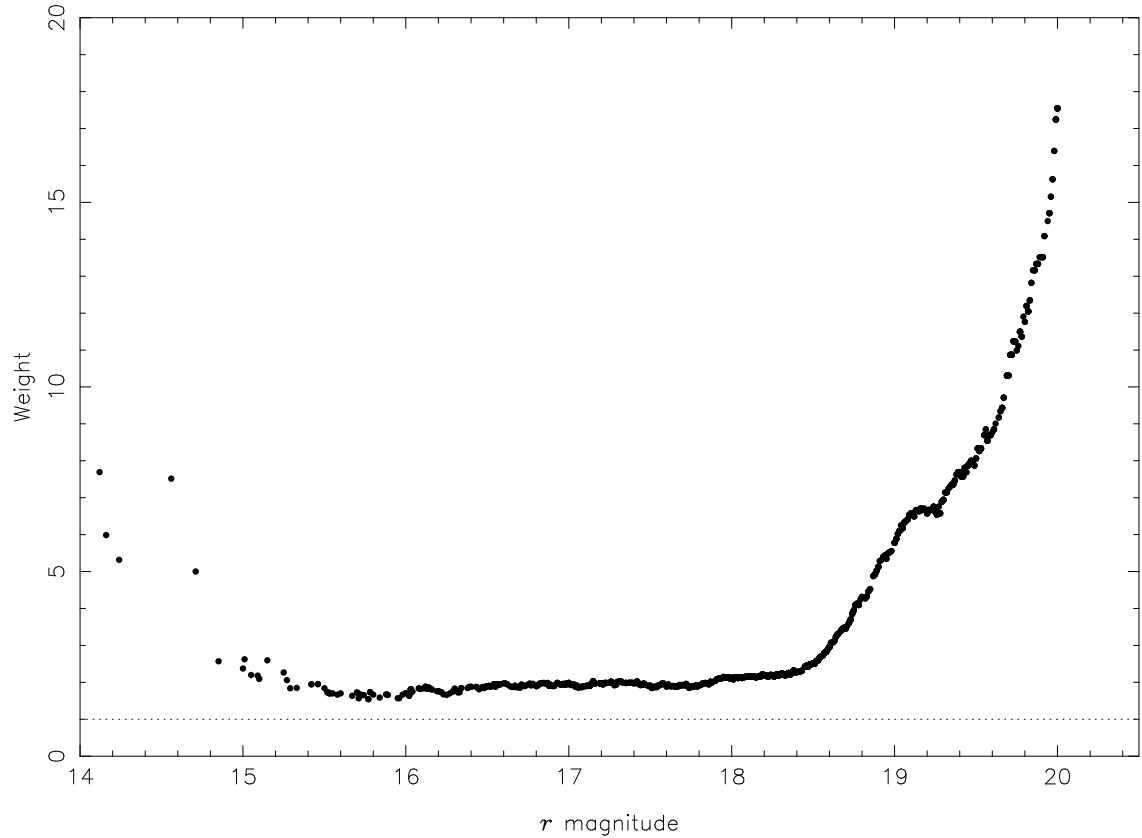}{4.3 in}{-90}{62}{62}{-234}{360}
\caption[]
{Selection weights as a function of $r$ magnitude.  The weights, even
for bright galaxies, do not reach unity because we sparsely sampled
our survey area.  Since scattered light from the very brightest galaxies
with $r \le 15^m$
would contaminate adjacent spectra on the CCD, we purposely avoided
observing such galaxies, and thus the weights actually increase
for $r \le 15^m$.}
\label{figures:weights}
\end{figure}

\newpage
\clearpage

\begin{figure}
\plotfiddle{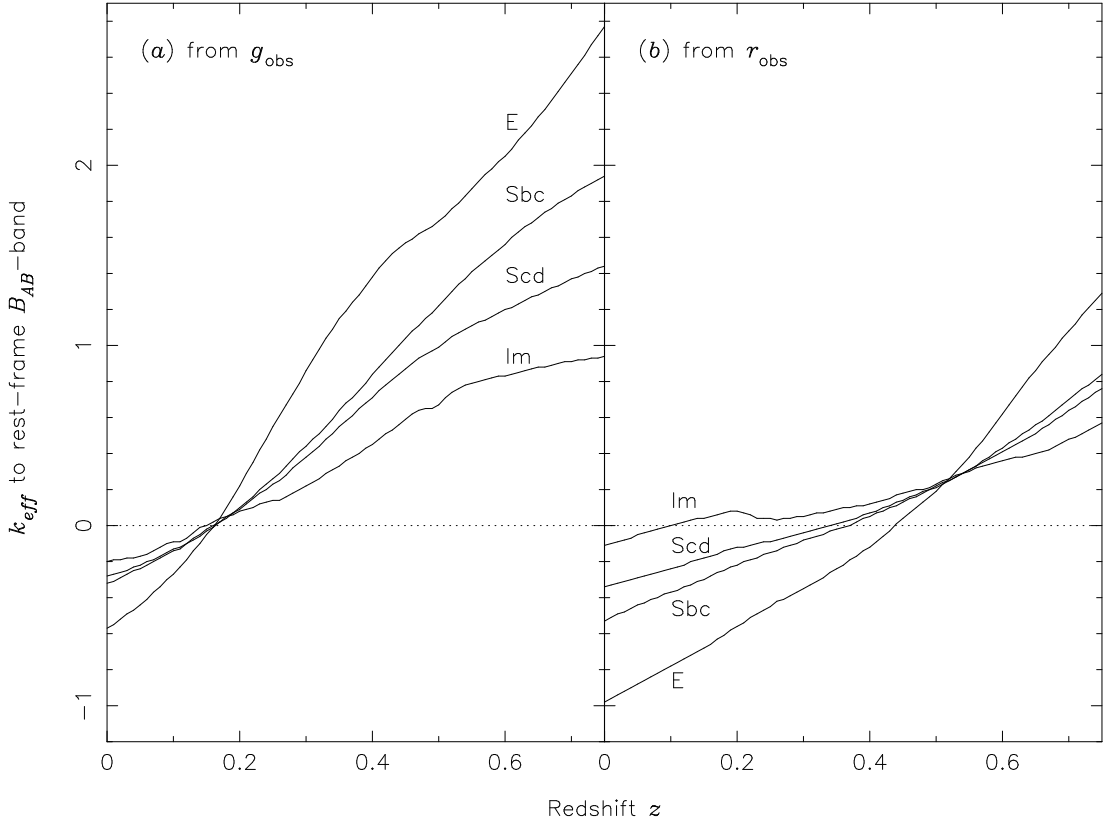}{4.3 in}{-90}{62}{62}{-234}{360}
\caption[]
{Effective $k$-corrections from the observed ($a$) $g$ and ($b$) $r$ bands to
the rest-frame $B_{AB}$ band.  The effective $k$-correction, $k_{eff}$, is
the traditional $k$-correction with the bandwidth stretching term
removed and the rest-frame color correction to the $B_{AB}$ band added.
It thus incorporates all of the spectrum-dependent corrections required
to transform from the observed band to the rest-frame $B_{AB}$ band.
The four curves in each panel labeled E, Sbc, Scd, and Im are the
four spectral types from CCW.  By transforming
from the observed $g$ band for $z \lesssim 0.3$ and from the observed
$r$ band for $0.3 \lesssim z \lesssim 0.5$, the correction is always less
than $\pm0.6^m$.}
\label{figures:keff}
\end{figure}

\newpage
\clearpage

\begin{figure}
\plotfiddle{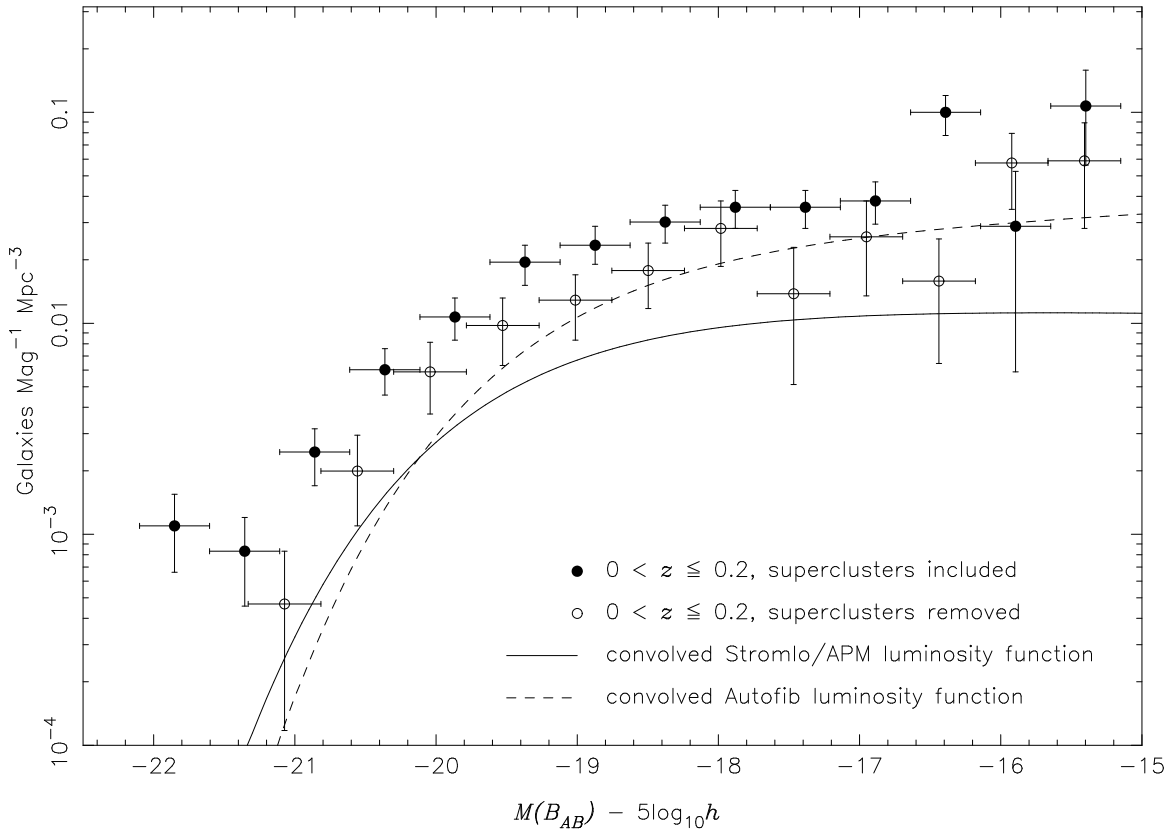}{4.3 in}{-90}{62}{62}{-234}{360}
\caption[]
{$B_{AB}$-band LF for galaxies with $0 < z \le 0.2$.
The unfilled circles show the LF with the two superclusters
at $z \approx 0.07$ and $z \approx 0.11$ removed.  The filled circles
show the LF with the superclusters included.  
The solid and dashed lines give the Schechter function
fits to the Stromlo/APM and Autofib local LFs, respectively,
convolved with a Gaussian of dispersion $0.25^m$ to account the
random magnitude errors in our survey.  (When reproducing the LFs
of other surveys in subsequent figures, we will always convolve
the published LFs with a Gaussian of dispersion $0.25^m$.)
While all the $B_{AB}$-band local LF's have similar shapes, there are
substantial differences in the normalization.}
\label{figures:local_lf}
\end{figure}

\newpage
\clearpage

\begin{figure}
\plotfiddle{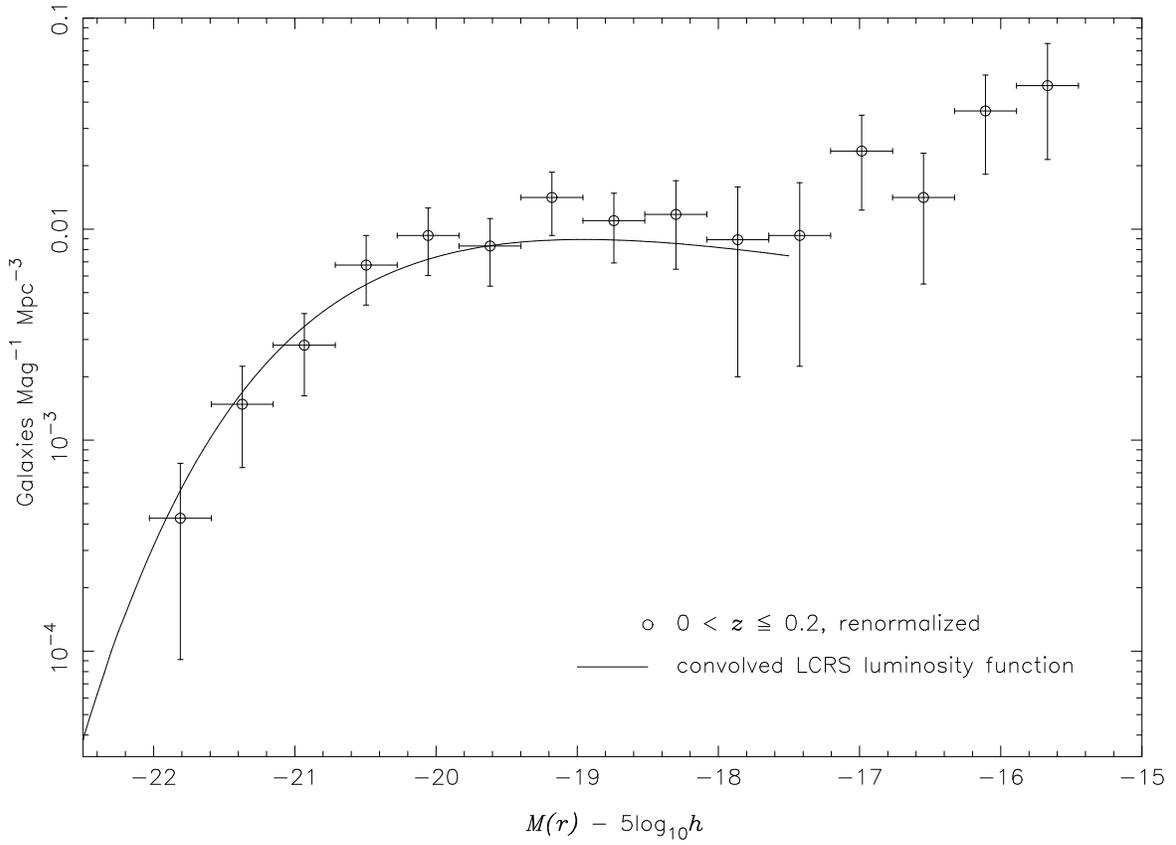}{4.3 in}{-90}{62}{62}{-234}{360}
\caption[]
{$r$-band LF for galaxies with $0 < z \le 0.2$.
The unfilled circles show our $r$-band local LF,
normalized to the counts of Weir et al. (1995).  The solid line
gives Schechter function fit to the local LF measured
in the LCRS.}
\label{figures:local_lf_r}
\end{figure}

\newpage
\clearpage

\begin{figure}
\plotfiddle{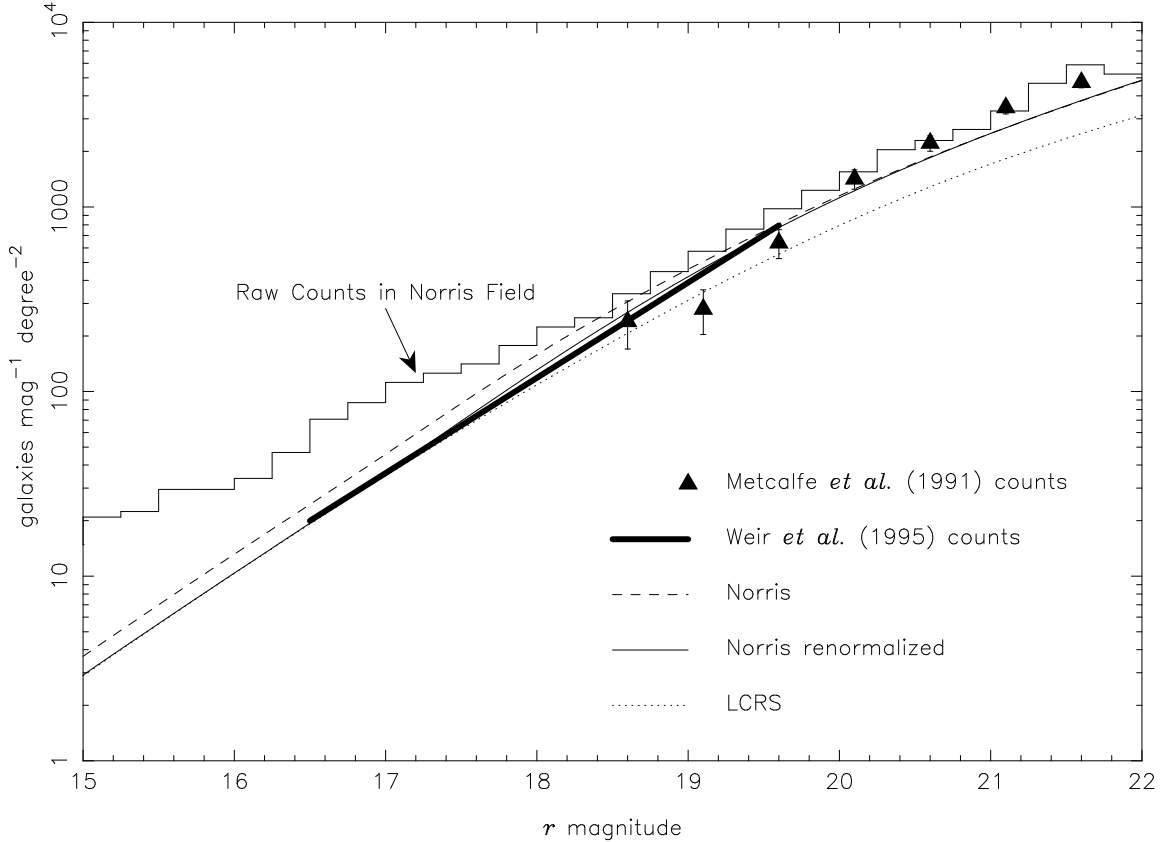}{4.3 in}{-90}{62}{62}{-234}{360}
\caption[]
{$r$-band differential galaxy counts.  The counts in POSS-II field covering
Corona Borealis, the POSS-II counts of Weir et al. (1995), and the
CCD counts of Metcalfe et al. (1991) are given by the histogram,
the thick solid line, and the triangles, respectively.  We also
plot the predicted differential number counts, computed using Equation
\ref{equations:diff_counts}, for three different LFs.  The dashed line
is based on the local LF normalized to the counts in the Corona Borealis
field with the supercluster removed and on the LF computed for
$0.2 < z \le 0.5$.  Normalizing our local LF to the counts of
Weir et al. (1995) for $16.5^m \le r \le 17.0^m$ and including the
evolution of the LF for $z > 0.2$, we obtain the predicted counts
marked by the solid line, which agree well with the observed counts
for $r \le 20.0^m$.  For comparison, we also plot the predicted
counts based on the LCRS local LF and not including evolution of the
LF beyond $z = 0.2$.}
\label{figures:r_counts}
\end{figure}

\newpage
\clearpage

\begin{figure}
\plotfiddle{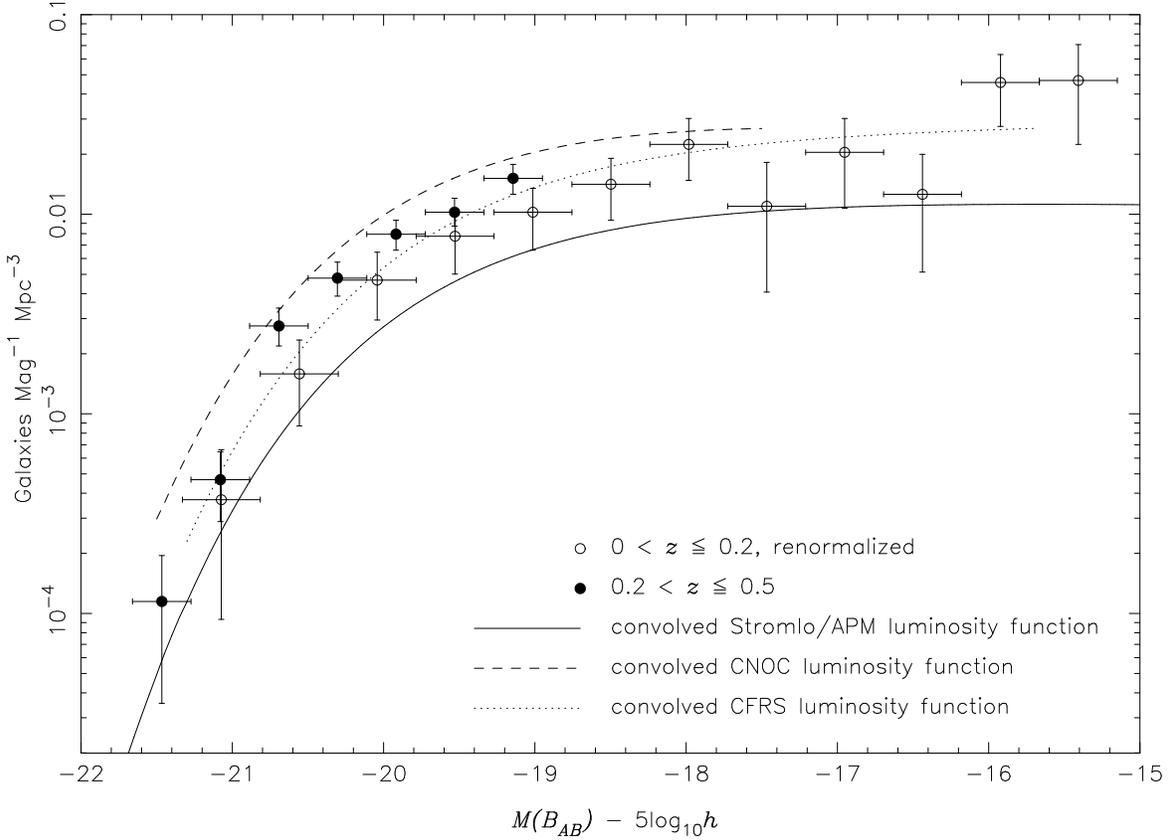}{4.3 in}{-90}{62}{62}{-234}{360}
\caption[]
{LFs for galaxies between $z = 0$ and $z = 0.5$.
The filled circles show the LF for $0.2 < z \le 0.5$,
while the unfilled circles show the LF for $0 < z \le 0.2$
normalized to the Weir et al. (1995) counts.  The solid line is the convolved
Schechter function fit to the local Stromlo/APM LF.  The dashed and
dotted
lines give the convolved Schechter function fits to the LFs
for $0.2 < z < 0.6$ and $0.2 < z < 0.5$ for CNOC and CFRS, respectively.
Evolution of the LF is apparent.}
\label{figures:field_lf}
\end{figure}

\newpage
\clearpage

\begin{figure}
\plotfiddle{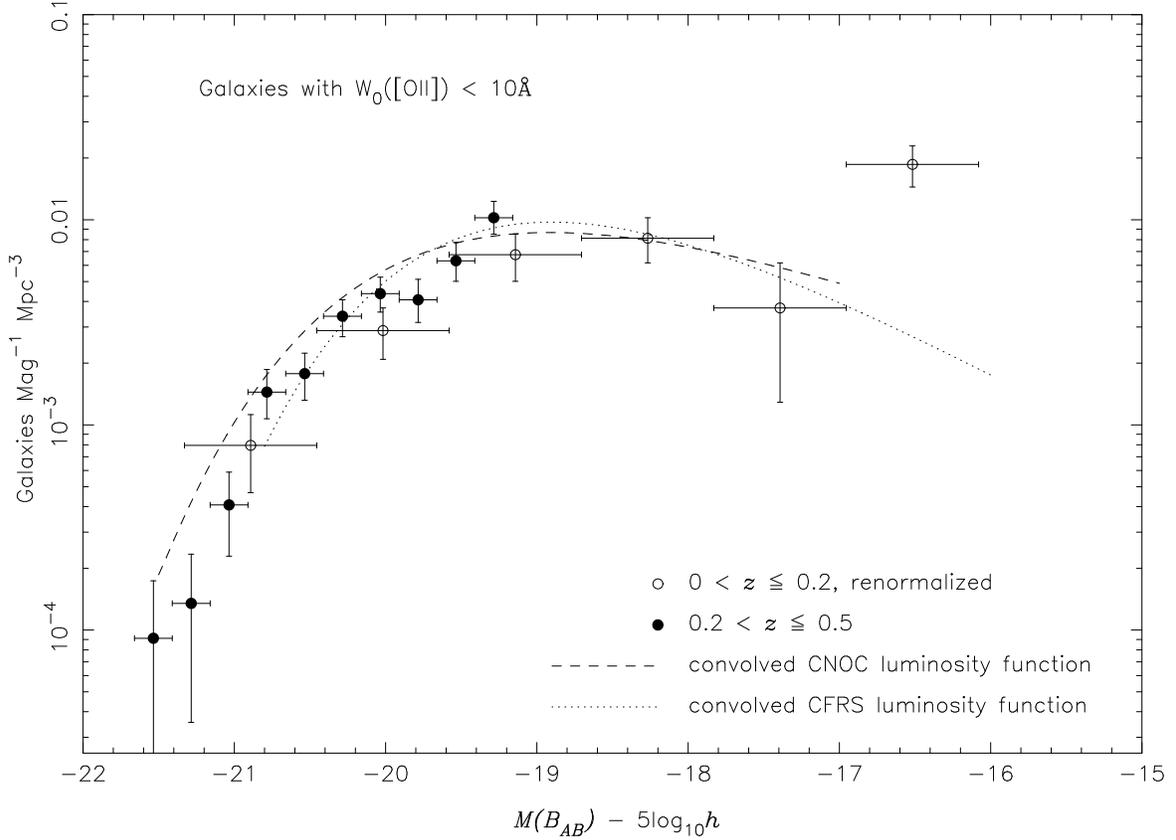}{4.3 in}{-90}{62}{62}{-234}{360}
\caption[]
{LF for galaxies with $W_0$([\ion{O}{2}]) $<$ 10\AA\ 
between $z = 0$ and $z = 0.5$.
The filled circles are the LF of quiescent galaxies for
$0.2 < z \le 0.5$; the unfilled circles are the local LF
for quiescent galaxies with the normalization reduced by 21\%.  
The dashed and dotted lines show the LF of 
corresponding color-selected galaxies (i.e., colors redder than those of
a CCW Sbc galaxy) in the redshift intervals $0.2 < z < 0.6$
and $0.2 < z < 0.5$ from CNOC and CFRS, respectively.  There is
no sign of evolution of the population of non-star-forming galaxies
from $z = 0$ to $z = 0.5$.} 
\label{figures:no_o2_lf}
\end{figure}

\newpage
\clearpage

\begin{figure}
\plotfiddle{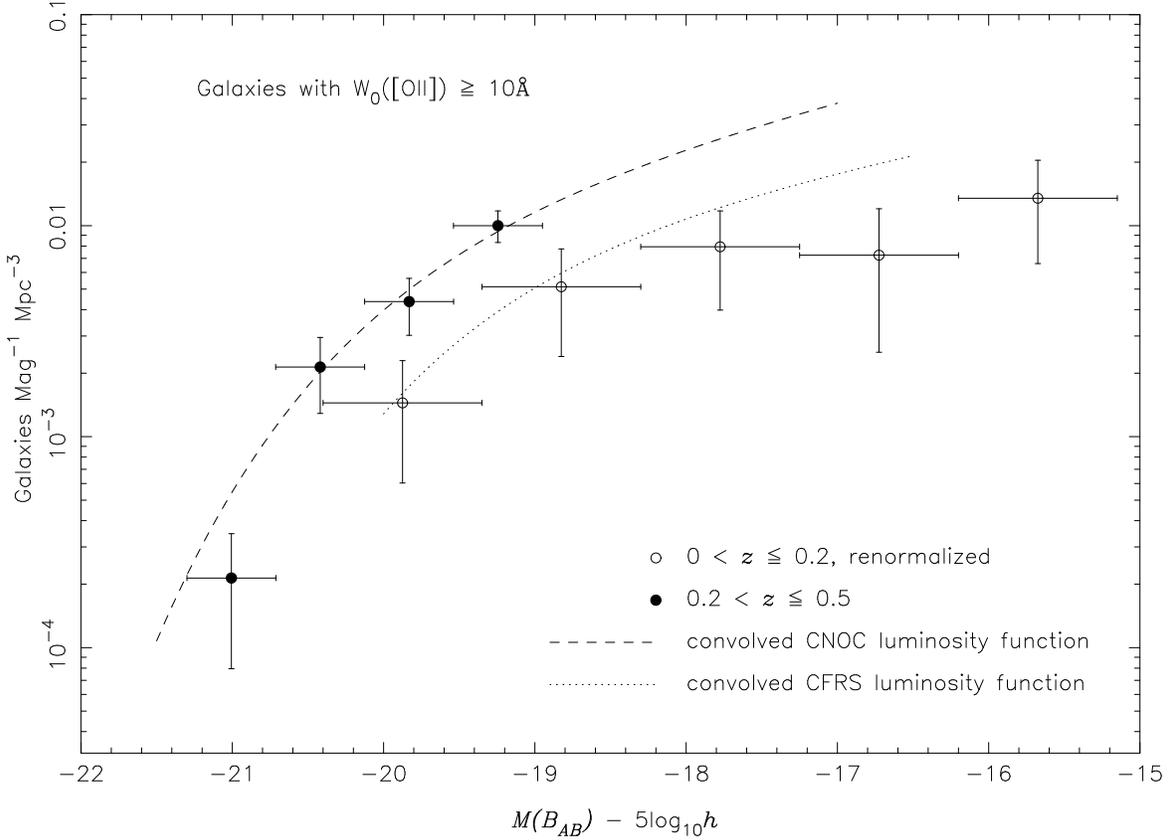}{4.3 in}{-90}{62}{62}{-234}{360}
\caption[]
{LF for galaxies with $W_0$([\ion{O}{2}]) $>$ 10\AA\ 
between $z = 0$ and $z = 0.5$.
The filled circles are the LF of star-forming galaxies for
$0.2 < z \le 0.5$; the unfilled circles are the local luminosity
function for star-forming galaxies with the normalization reduced
by 21\%.
The dashed and dotted lines show the LF of 
corresponding color-selected galaxies (i.e., colors bluer than those of
a CCW Sbc galaxy) in the redshift intervals $0.2 < z < 0.6$
and $0.2 < z < 0.5$ from CNOC and CFRS, respectively.  
The population of star-forming
galaxies has evolved rapidly from $z = 0$ to $z = 0.5$.}
\label{figures:o2_lf}
\end{figure}

\newpage
\clearpage

\begin{figure}
\plotfiddle{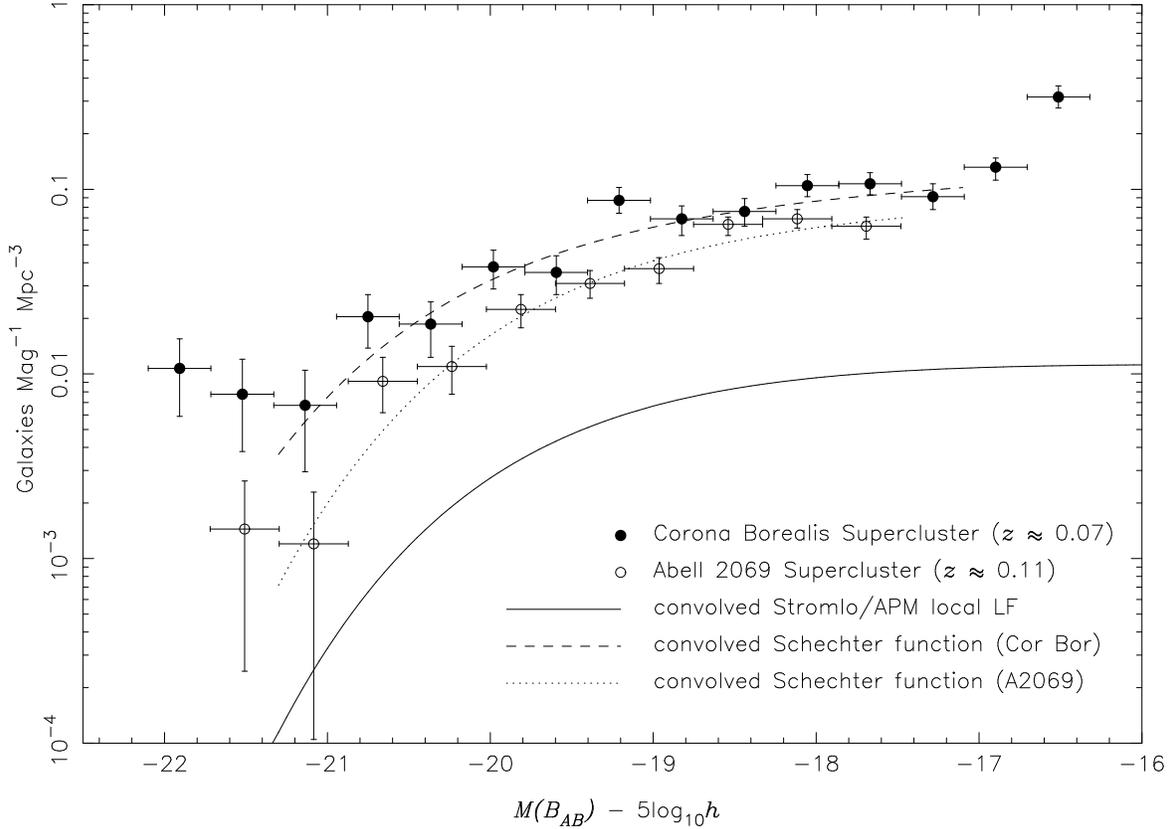}{4.3 in}{-90}{62}{62}{-234}{360}
\caption[]
{LFs of the Corona Borealis supercluster (filled circles)
and the background A2069 supercluster (unfilled circles).  
The convolved Schechter function fits to the two supercluster luminosity
functions ($M(B_{AB}) > -21.3^m$)
are given by the dashed (Corona Borealis) and dotted (A2069)
lines.  Brighter than $M(B_{AB}) \sim -21^m$, the superclusters have
excess galaxies relative to the Schechter function fit.  In addition,
there is a hint that the LF of the Corona Borealis supercluster has
a sharp upturn fainter than $M(B_{AB}) \sim -17^m$.
The local LF from the Stromlo/APM
survey is also plotted for comparison.}
\label{figures:clusters_lf}
\end{figure}

\newpage
\clearpage

\begin{figure}
\plotfiddle{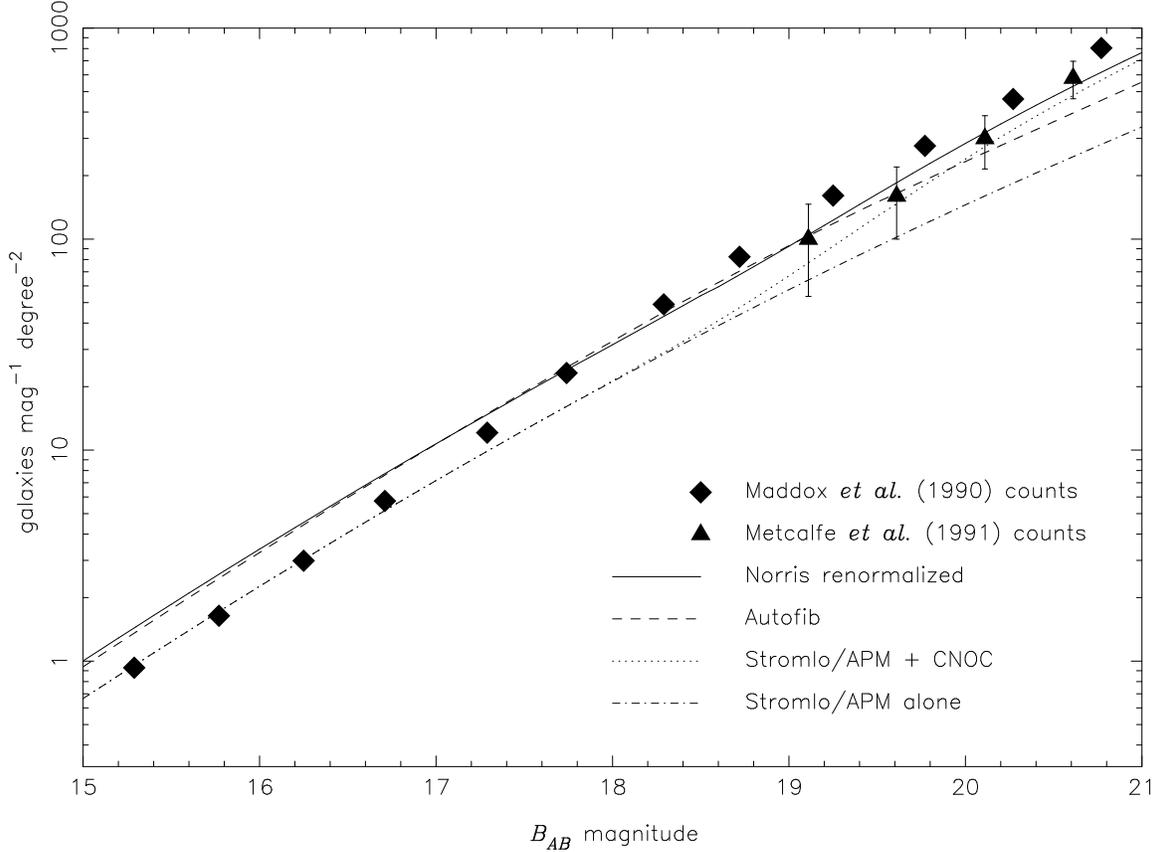}{4.3 in}{-90}{62}{62}{-234}{360}
\caption[]
{Observed and predicted differential counts in the $B_{AB}$ band.  The diamonds
are APM counts from Maddox et al. (1990), while the triangles are CCD-based
counts from Metcalfe et al. (1991).  We plot the predicted counts,
using Equation \ref{equations:diff_counts}, for several different observed
LFs.  The solid line gives the predicted counts for the evolving luminosity
function measured in our survey (with the local LF normalized to the
$r$-band counts).  Similarly, the dashed line is the prediction from the
Autofib survey, the dotted line the prediction using the Stromlo/APM local
LF and the CNOC LF for $z > 0.2$, and the dot-dashed line the prediction
using just the Stromlo/APM LF alone.  Our predictions and those based
on the Autofib LF match the counts well for $18^m \lesssim B_{AB} \lesssim
21^m$, whereas the prediction which incorporates the Stromlo/APM local LF
and the CNOC LF for $z > 0.2$ matches the counts for $B_{AB} \lesssim 17^m$
and for $B_{AB} \gtrsim 19^m$, where most of the galaxies are at $z > 0.2$.}
\label{figures:B_counts}
\end{figure}

\newpage
\clearpage

\begin{figure}
\plotfiddle{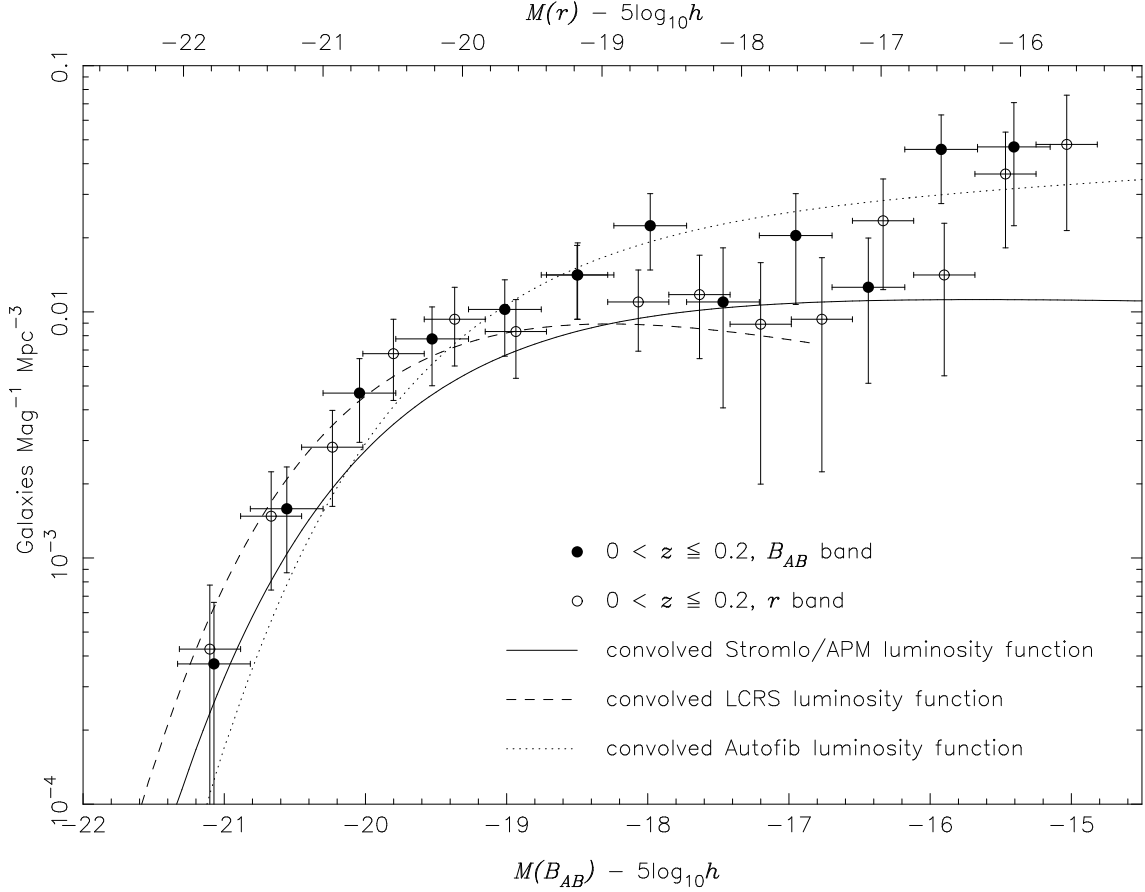}{4.3 in}{-90}{62}{62}{-234}{360}
\caption[]
{Assorted $B_{AB}$- and $r$-band local LFs.  The
$B_{AB}$-band local LFs are plotted with respect to the bottom axis,
while the $r$-band local LFs are plotted with respect to the top
axis.  The two axes are offset by $B_{AB}-r = 0.72^m$, which is the
median color we measure for field galaxies with $z < 0.2$.  The
filled circles are the $B_{AB}$-band local LF, and the unfilled circles
are the $r$-band local LF.  The normalizations of both of our local
LFs have been reduced by 21\%.
The $B_{AB}$-band local LFs from the Stromlo/APM survey and the
Autofib survey are plotted with the solid and dotted lines, respectively.
The dashed line represents the local LF from the LCRS survey.
After applying the color offset of $B_{AB}-r = 0.72^m$, our $B_{AB}$-
and $r$-band local LFs agree well.  However, in contrast to Lin
et al. (1996b), we do not conclude that the LCRS $r$-band local LF
matches the Stromlo/APM LF.}
\label{figures:B_and_r_lf}
\end{figure}

\newpage
\clearpage

\begin{figure}
\plotfiddle{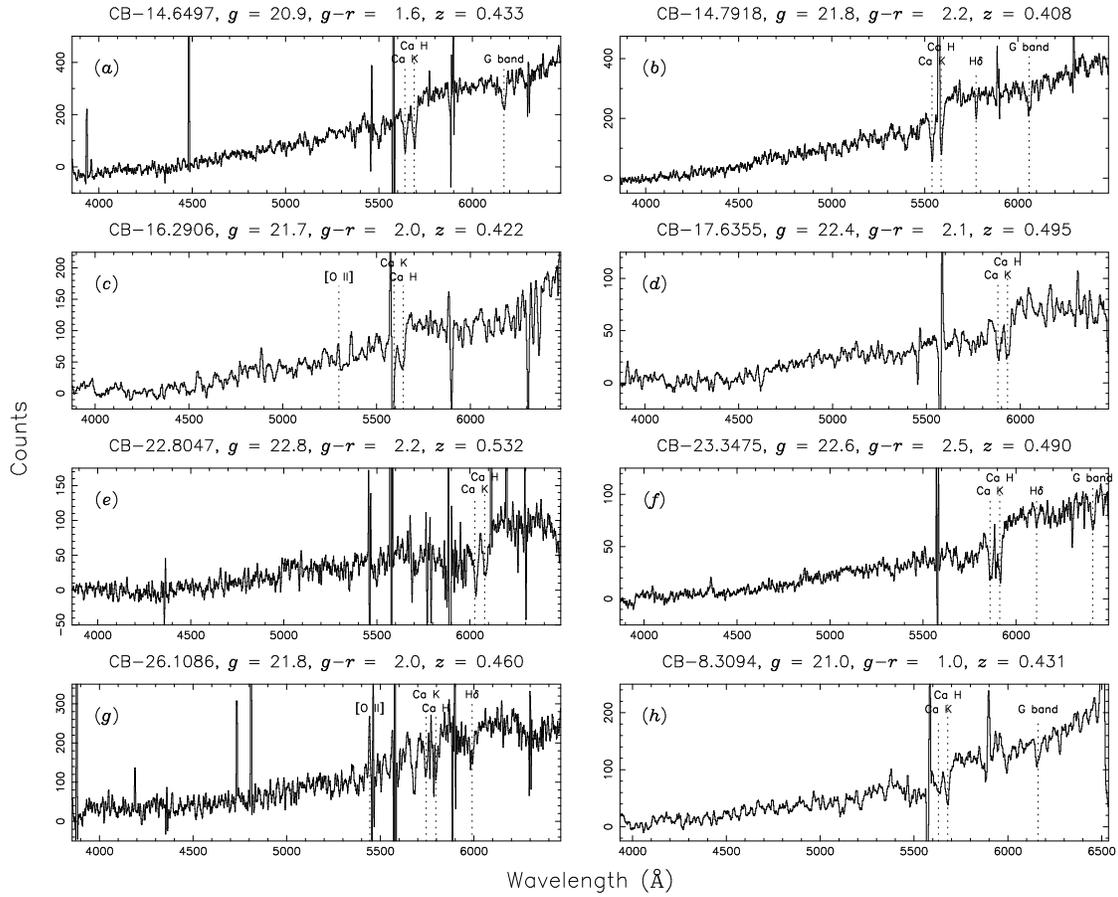}{4.3 in}{-90}{62}{62}{-234}{360}
\caption[]
{A representative sample of absorption-line galaxies at $0.4 < z < 0.6$
illustrating the strength of Ca H, Ca K, and the 4000\AA\ break.}
\label{figures:ab_line_objs}
\end{figure}

\newpage
\clearpage

\begin{figure}
\plotfiddle{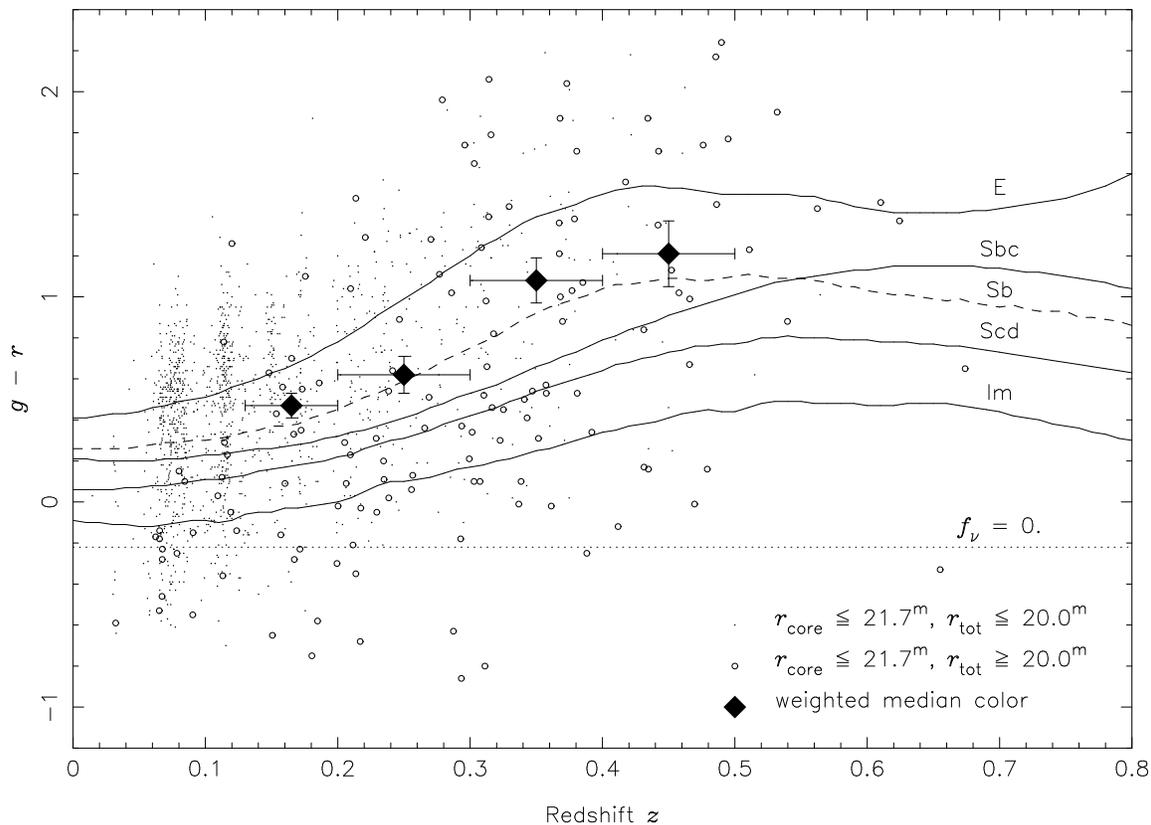}{4.3 in}{-90}{62}{62}{-234}{360}
\caption[]
{Median observed $g-r$ color as a function of redshift.  We plot
the observed $g-r$ color as a function of redshift along with the
tracks of six representative model spectra.  The bluest track is
that of flat-spectrum galaxy, $f_\nu = 0.$  The other five tracks
are those typical of the Hubble types E, Sb, Sbc, Scd, and Im.  The large,
filled triangles mark the weighted median $g-r$ color in the redshift intervals
$0.13 < z < 0.2$, $0.2 < z < 0.3$, $0.3 < z < 0.4$, and $0.4 < z < 0.5$.
The Sb track (the dotted line, from Bruzual \& Charlot 1993) matches the
observed median galaxy colors to $z \sim 0.5$.}
\label{figures:obs_g-r}
\end{figure}

\newpage
\clearpage

\begin{figure}
\plotfiddle{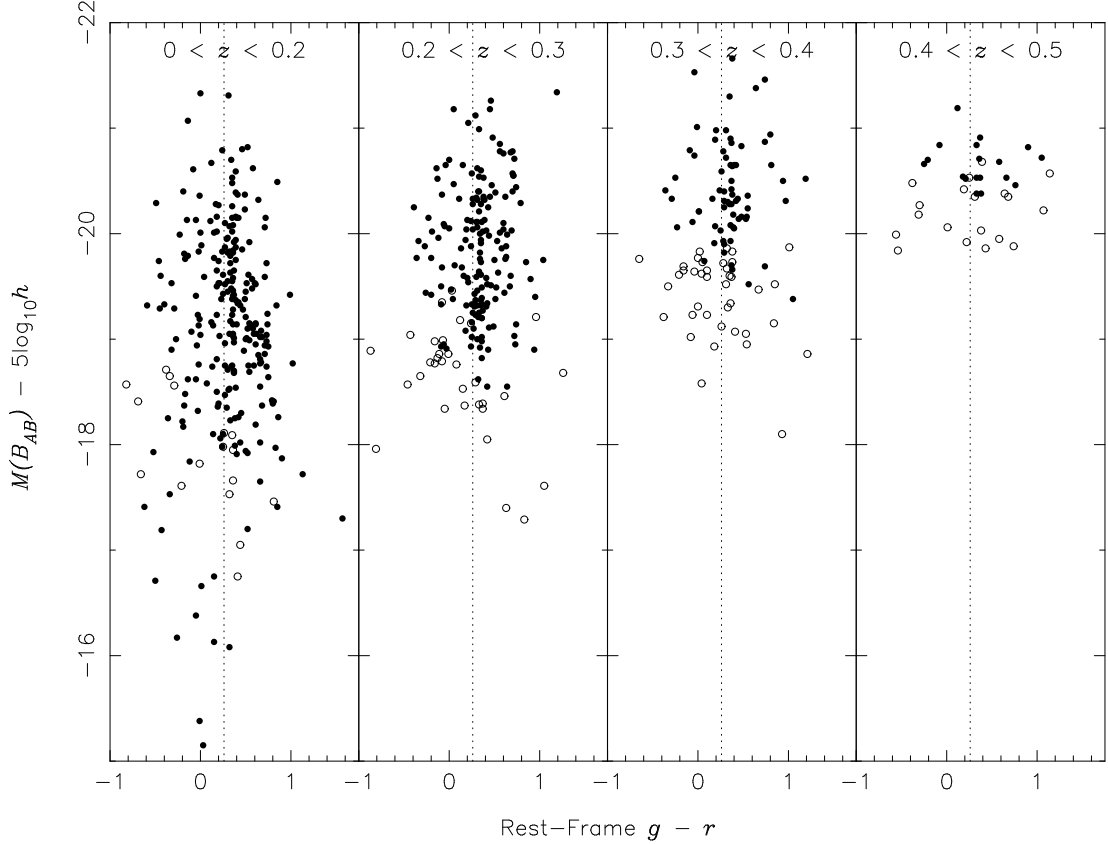}{4.3 in}{-90}{62}{62}{-234}{360}
\caption[]
{Absolute $B_{AB}$-band magnitude versus rest-frame $g-r$ colors for
our survey galaxies, divided into four redshift intervals,
$0 < z < 0.2$ (with the superclusters removed), $0.2 < z < 0.3$,
$0.3 < z < 0.4$, and $0.4 < z < 0.5$.  Galaxies with $r \le 20.0^m$
are denoted by filled circles, and those with $r > 20.0^m$ with
unfilled circles.
For $z < 0.2$, the well-known
color-magnitude relationship, i.e., that brighter galaxies are redder,
is evident.  However, at higher redshifts, the color-magnitude relationship
disappears as the population of blue galaxies evolves and becomes
intrinsically brighter.}
\label{figures:rest-frame_g-r}
\end{figure}

\newpage
\clearpage

\begin{figure}
\plotfiddle{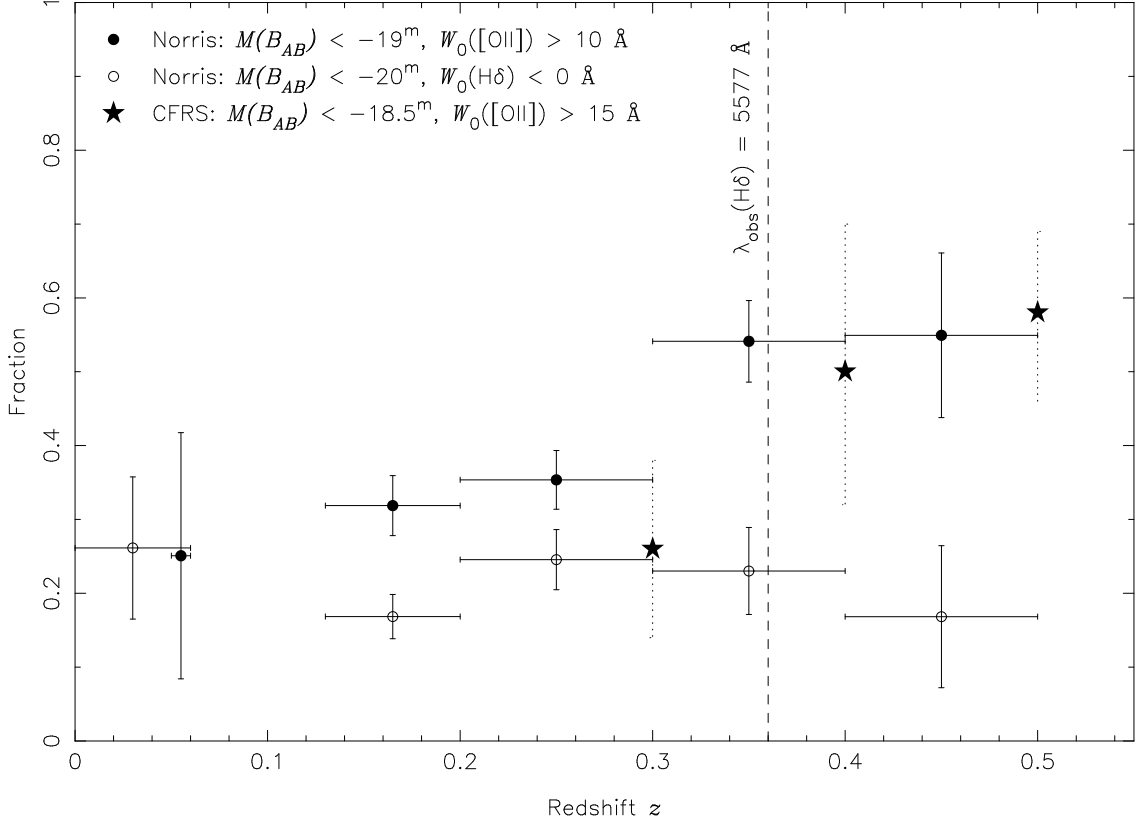}{4.3 in}{-90}{62}{62}{-234}{360}
\caption[]
{Fraction of galaxies with [\ion{O}{2}] $\lambda$3727 emission
(filled circles)
and H$\delta$ $\lambda$4101 absorption (unfilled circles)
as a function of redshift.  
The superclusters are excluded from the analysis.  The dashed
line indicates the redshift at which the observed wavelength
of H$\delta$ is 5577 \AA; redward of this wavelength the increasingly
bright sky makes measurement of the strength of H$\delta$
more uncertain.
The filled stars give the 
fraction of galaxies in the CFRS with $M(B_{AB}) < -18.5^m$ and
[\ion{O}{2}] $>$ 15\AA\ (Hammer et al. 1996).
As [\ion{O}{2}] is an indicator of ongoing
star formation, the increasing fraction of galaxies exhibiting
[\ion{O}{2}] with redshift is evidence for an increase in the rate
of star formation with redshift.  However, the fraction of galaxies
with H$\delta$ absorption does not change with redshift, which argues
against the star-formation activity in intermediate-redshift galaxies
occurring predominantly in discrete bursts.}
\label{figures:o2_and_hd}
\end{figure}

\newpage
\clearpage

\begin{figure}
\plotfiddle{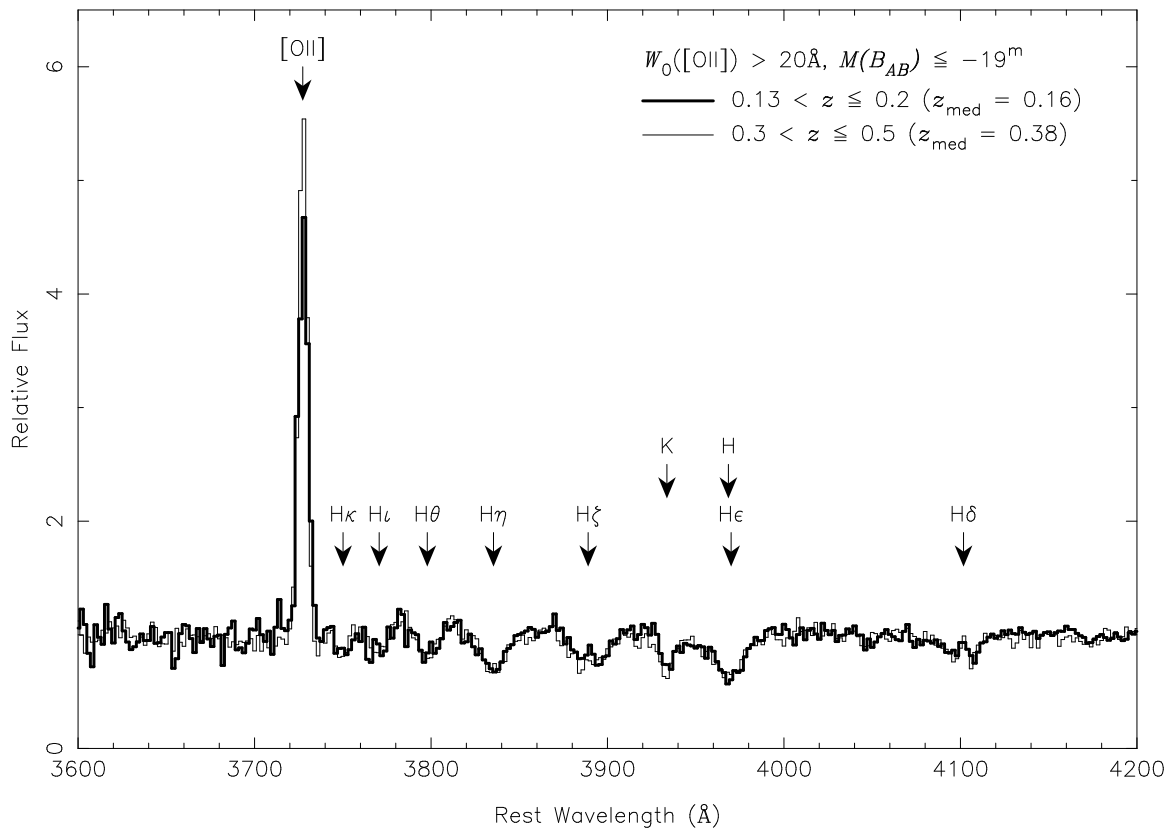}{4.3 in}{-90}{62}{62}{-234}{360}
\caption[]
{Composite spectra of galaxies with $W_0$[\ion{O}{2}] $>$ 20\AA\ and
$M_{B_{AB}} \le -19^m$ for two redshift ranges, $0.13 < z \le 0.2$ 
(thick line) and $0.3 < z \le 0.5$ (thin line). 
The high redshift composite spectrum has stronger
[\ion{O}{2}], but the spectra are otherwise identical.  In particular,
the strength of H$\delta$ has not changed with redshift.}
\label{figures:coadd}
\end{figure}

\newpage
\clearpage

\begin{figure}
\plotfiddle{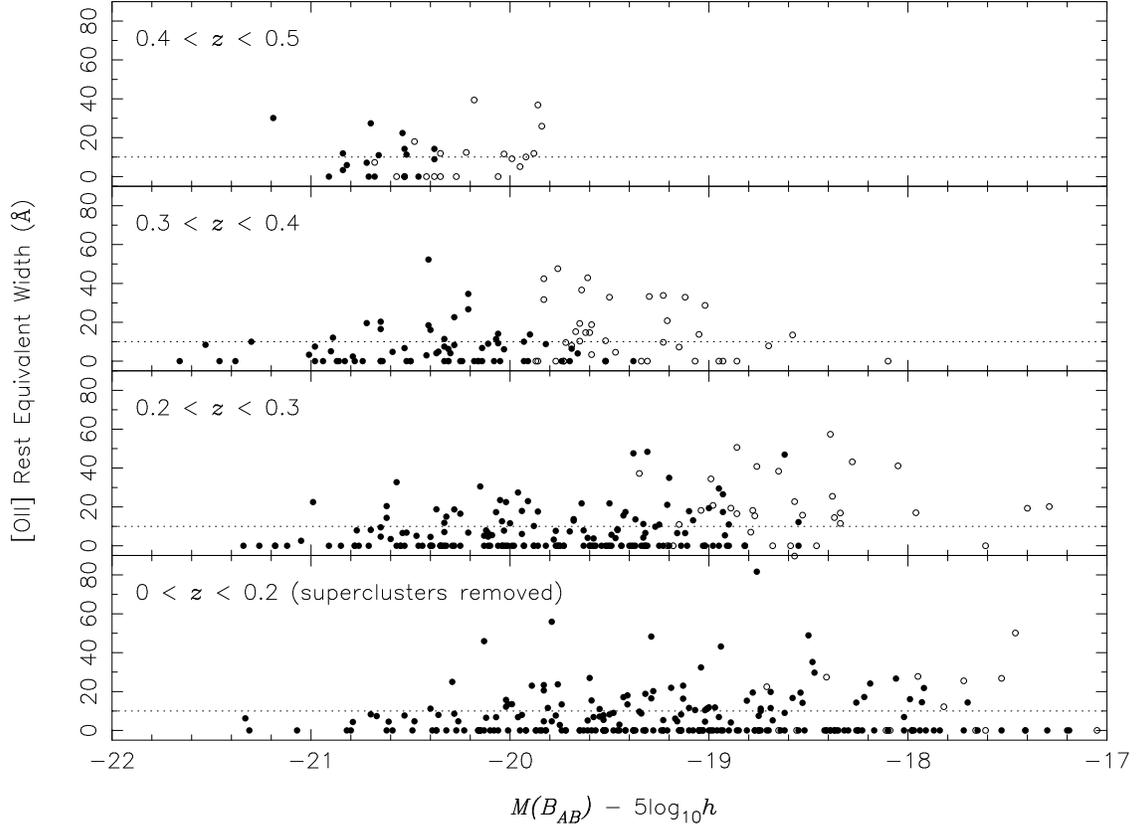}{4.3 in}{-90}{62}{62}{-234}{360}
\caption[]
{Rest equivalent widths of [\ion{O}{2}] as a function of absolute
$B_{AB}$-band magnitude in four redshift intervals, $0 < z < 0.2$ (with
the superclusters removed), $0.2 < z < 0.3$, $0.3 < z < 0.4$, and
$0.4 < z < 0.6$.  The filled circles have $r \le 20^m$, and the
unfilled circles have $r > 20^m$.  The dotted line marks an [\ion{O}{2}]
equivalent width of 10\AA.  This figure illustrates two important
points.  The range of star formation rates, as measured by the
[\ion{O}{2}] equivalent widths, at $z \sim 0.5$ is similar to
that at low redshift.  However, the luminosities
of galaxies undergoing rapid star formation increases progressively
from low redshift, where ongoing star formation is mainly restricted
to sub-$L^\ast$ galaxies, to high redshift, where even some super-$L^\ast$
galaxies are rapidly forming stars.}
\label{figures:o2_ew}
\end{figure}

\end{document}